\documentclass[prb,twocolumn,showpacs,amsmath,amssymb,superscriptaddress]{revtex4}

\usepackage{graphicx}
\usepackage{color}
\usepackage{amsmath,amssymb,mathrsfs,ulem}
\usepackage{bbm}
\usepackage[dvipsnames]{xcolor}
\usepackage{tikz}
\usepackage{pgfplots}
\usetikzlibrary{calc,arrows,decorations.pathmorphing,backgrounds,positioning,fit,petri,shapes,intersections,external}

\newcommand{\ket}[1]{\left|#1\right\rangle}
\newcommand{\bra}[1]{\left\langle#1\right|}
\newcommand{\dunderline}[1]{\underline{\underline{#1}}}

\date{\today}                  

\begin{document}

\title{Flavor fluctuations in $3$-level quantum dots: Generic $SU(3)$-Kondo fixed point
  in equilibrium and non-Kondo fixed points in nonequilibrium}

\author{Carsten J. Lindner}
\affiliation{Institut f\"ur Theorie der Statistischen Physik, RWTH Aachen, 
52056 Aachen, Germany and JARA - Fundamentals of Future Information Technology}
\author{Fabian B. Kugler}
\affiliation{Physics Department, Arnold Sommerfeld Center for Theoretical Physics, and Center
for NanoScience,Ludwigs-Maximilians-Universit\"at M\"unchen, Theresienstr. 37, 80333 Munich, Germany}
\author{Herbert Schoeller}
\email[Email: ]{schoeller@physik.rwth-aachen.de}
\affiliation{Institut f\"ur Theorie der Statistischen Physik, RWTH Aachen, 
52056 Aachen, Germany and JARA - Fundamentals of Future Information Technology}
\author{Jan von Delft}
\affiliation{Physics Department, Arnold Sommerfeld Center for Theoretical Physics, and Center
for NanoScience,Ludwigs-Maximilians-Universit\"at M\"unchen, Theresienstr. 37, 80333 Munich, Germany}

\begin{abstract}
We study a $3$-level quantum dot in the singly occupied cotunneling regime coupled 
via a generic tunneling matrix to several multi-channel leads in equilibrium or nonequilibrium. 
Denoting the three possible states of the quantum dot by the quark flavors up ($u$), down ($d$) 
and strange ($s$), we derive an effective model where also each reservoir has three flavors labelled by 
$u$, $d$ and $s$ with an effective density of states polarized w.r.t. an 
eight-dimensional $F$-spin corresponding to the eight generators of $SU(3)$.
In {\it equilibrium} we perform a standard poor man scaling analysis 
and show that tunneling via virtual intermediate states induces flavor fluctuations 
on the dot which become $SU(3)$-symmetric at a characteristic and exponentially small low-energy 
scale $T_K$. Close to $T_K$ the system is described by a 
single isotropic Kondo coupling $J>0$ diverging at $T_K$. Using the numerical
renormalization group (NRG) we study in detail the linear conductance and 
confirm the $SU(3)$-symmetric Kondo fixed point with universal conductance 
$G=3\sin^2(\pi/3){e^2\over h}=2.25{e^2\over h}$ for 
various tunneling setups by tuning the level spacings on the dot. We also identify regions 
of the level positions where the $SU(2)$-Kondo fixed point is obtained and find a rather complex 
dependence of the various Kondo temperatures as function of the gate voltage and the 
tunneling couplings. In contrast to the equilibrium case, we find in {\it nonequilibrium} that
the fixed point model is {\it not} $SU(3)$-symmetric but characterized by rotated $F$-spins for 
each reservoir with total vanishing sum.
At large voltage we analyse the $F$-spin magnetization and the current in golden rule as function
of a longitudinal ($h_z$) and perpendicular ($h_\perp$) magnetic field for the isospin and the 
level spacing $\Delta$ to the strange quark. As a smoking gun to detect the nonequilibrium fixed point
we find that the curve of zero $F$-spin magnetization in $(h_z,h_\perp,\Delta)$-space is a circle when
projected onto the $(h_z,h_\perp)$ plane. We propose that our findings can be generalized to the
case of quantum dots with an arbitrary number $N$ of levels. 

\end{abstract}


\maketitle

\section{Introduction}
\label{sec:introduction}

Over the last three decades transport properties of correlated quantum dots have gained an enormous
interest in many experimental and theoretical research activities in condensed matter physics. As
artificial atoms they allow for a controlled study of interesting phenomena playing a central role
in many different fields of applied and fundamental research in nanoelectronics, spintronics, 
quantum information processing, dissipative quantum mechanics, and many-body physics and 
nonequilibrium phenomena in correlated systems, see e.g. Refs.~\onlinecite{review_qi,andergassen_etal} for 
reviews. Of particular interest is the cotunneling or Coulomb blockade regime of quantum dots with strong
charging energy, where the charge is fixed and only the spin and orbital degrees of freedom can 
fluctuate by second-order tunneling processes via virtual intermediate states. In this regime 
effective models can be derived which are equivalent to Kondo models well-known from solid state
physics \cite{hewson_97}, see e.g. Ref.~\onlinecite{glazman_pustilnik_05} for a review of the 
Kondo effect in quantum dots. The standard model is the $SU(2)$-Kondo model, where a local 
spin-${1\over 2}$ is coupled via an isotropic exchange coupling to the spins of two large reservoirs.
Below a characteristic low energy scale, called the Kondo temperature $T_K$, the local spin is
completely screened and the remaining potential scattering leads to resonant transport through the
system with universal conductance $2{e^2\over h}$. This Kondo effect has been theoretically predicted 
for quantum dots \cite{kondo_theo} and has been experimentally observed \cite{kondo_exp}. 
After this discovery the research for Kondo physics in quantum dots has gained an enormous interest
and further realizations have been proposed and observed, like e.g. the realization of higher spin values
\cite{higher_spin_kondo}, singlett-triplett fluctuations \cite{ST_kondo}, 
non-Fermi liquid behaviour in $2$-channel realizations \cite{2_channel_kondo}, and 
the $SU(4)$-Kondo effect \cite{SU4_kondo}. Recently, also the realization of $SU(N)$-Kondo physics for
arbitrary $N$ has been proposed in coupled quantum dots \cite{coupled_qd,Moca2012,lopez_etal_13}.

The enormous variety of possible realizations of Kondo physics raises the question what happens in the
generic case when a quantum dot in the regime of fixed charge with $N_\text{dot}\ge 1$ electrons and 
$N\ge 2$ levels is coupled via a generic tunneling matrix to several multi-channel reservoirs. Even for
the simplest case $N_\text{dot}=1$ and $N=2$, this issue is nontrivial since the quantum number $l=1,2$ 
labelling the two dot levels is in general a non-conserved quantity in tunneling, like e.g. 
for ferromagnetic leads \cite{koenig_etal}, orbital degrees of freedom \cite{boese_etal}, 
Aharonov-Bohm geometries \cite{kashcheyevs_etal_prb07}, and spin-orbit or
Dzyaloshinski-Moriya interactions \cite{paaske_etal_prb10,pletyukhov_schuricht_prb11}. 
In Ref.~\onlinecite{kashcheyevs_etal_prb07} it was shown via a singular value decomposition of
the total tunneling matrix (i.e. containing {\it all} reservoirs) that all these different cases can 
be mapped onto an effective model which is equivalent to the anisotropic spin-${1\over 2}$ Kondo model which
flows into the isotropic $SU(2)$-symmetric fixed point at low energies below the Kondo temperature. 
This explains why in all linear response transport calculations of quantum dot models with 
$N_\text{dot}=1$ and $N=2$, the Kondo effect with universal conductance is observed provided that
local effective magnetic fields are explicitly cancelled by external ones \cite{koenig_nrg}. 
However, this result is only valid in the linear response regime and for proportional 
couplings to all the reservoirs where the linear conductance can be related to
the equilibrium spectral density of the dot \cite{meir_wingreen_92}. To calculate the latter all reservoirs 
can be taken together to a single one and only the total tunneling matrix matters. 
However, when all reservoirs are coupled in a generic way to the dot or when
they are characterized by different temperatures or chemical potentials, the analysis of
Ref.~\onlinecite{kashcheyevs_etal_prb07} is no longer valid. This fact was emphasized in 
Ref.~\onlinecite{goettel_reininghaus_schoeller_15}, where it was shown that in a generic 
{\it nonequilibrium} situation, the proper effective model for $N_\text{dot}=1$ and $N=2$ is
a spin-valve model, where the spin polarizations of all reservoirs point in different directions, such
that at the low-energy fixed point their sum is equal to zero. This has the consequence that the
fixed point model in nonequilibrium is essentially not $SU(2)$-symmetric and new interesting
nonequilibrium fixed point models emerge with different non-Kondo like properties in the weak 
as well as in the strong coupling regime. Only in the equilibrium situation when all 
reservoirs are characterized by the same temperature and chemical potential, all reservoirs can be taken
together resulting in an unpolarized reservoir with $SU(2)$-symmetry at the fixed point. The
nonequilibrium properties at and away from the fixed point model have been studied for
large voltages above the Kondo temperature \cite{goettel_reininghaus_schoeller_15} and a 
smoking gun was identified in the nontrivial magnetic field dependence of the magnetization and the
transport current characterizing the fixed point model.

The proposals of new nonequilibrium fixed point models are of particular interest for the constant
effort to generalize well-established analytical and numerical methods for the study of equilibrium properties
of quantum impurity models \cite{costi_etal_94,hewson_97} to the nonequilibrium case. Recent developments of 
perturbative renormalization group methods \cite{rosch_etal,kehrein_etal,schoeller_rtrg,frg_noneq} 
have shown how the voltage dependence and the physics of cutoff scales by decay rates can be implemented 
\cite{hs_reininghaus_09} and how the time evolution into the stationary state can be 
calculated \cite{pletyukhov_etal_prl10}. Even in the strong coupling regime 
\cite{strong_coupling_RTRG,smirnov_grifoni_03} results in agreement with experiments 
\cite{strong_coupling_exp} were obtained, although the used methods are essentially perturbative
and not capable of describing the strong coupling regime in general. Therefore, numerically exact methods
are required for the description of quantum dot systems in nonequilibrium, like e.g. 
the time-dependent numerical renormalization group \cite{TD-NRG}, time-dependent density matrix
renormalization group \cite{TD-DMRG}, iterative stochastic path integrals \cite{ISPI}, and quantum 
Monte Carlo methods \cite{QMC}. Recently, a promosing thermofield approach has been suggested
by a combination of TD-NRG and TD-DMRG \cite{schwarz_etal_17} showing a good agreement with the
strong coupling results for the nonequilibrium Kondo model of 
Rfs.~\onlinecite{strong_coupling_RTRG,smirnov_grifoni_03,strong_coupling_exp}.

The aim of the present paper is to analyse the generic case $N_\text{dot}=1$ and arbitrary $N$ to
see how the results of Ref.~\onlinecite{goettel_reininghaus_schoeller_15} can be generalized to the
case $N>2$. In particular we will study the case $N=3$ and, starting from a generic tunneling matrix,
will show that an effective tunneling model can be derived where also the reservoirs are 
characterized by three flavors which we will conveniently label by the up ($u$),
down ($d$), and strange ($s$) quark flavors. 
The effective model in the cotunneling regime of a singly occupied
quantum dot can be described by flavor fluctuations and we will show by a poor man scaling analysis
that the low-energy fixed point model is indeed the $SU(3)$-symmetric Kondo model. This result is 
shown to hold also for arbitrary $N$ within the poor man scaling analysis and will be explicitly 
confirmed for $N=3$ by a numerically exact NRG analysis for the linear response conductance, 
similiar to Rfs.~\onlinecite{Moca2012,lopez_etal_13}. In addition to these references we will 
study the dependence of the $SU(3)$-Kondo temperature on the tunneling matrix elements and will show how
the $SU(3)$-symmetric point is obtained by a proper adjustment of the level spacings of the dot. 
Subsequently, we will analyse the nonequilibrium situation and generalize the spin-valve model of
Ref.~\onlinecite{goettel_reininghaus_schoeller_15} for $N=2$ to the case of three levels $N=3$. In 
this case a fixed point model arises where the reservoirs are characterized by eight-dimensional
$F$-spins corresponding to the eight generators of the $SU(3)$-group which cancel when all reservoirs
are taken together. For large voltages and two reservoirs we find that the nonequilibrium 
fixed point model has a characteristic dependence on the dot parameters for zero $F$-spin magnetization
on the dot providing a {\it smoking gun} for the detection of the fixed point. Thus we conclude that
the results of Ref.~\onlinecite{goettel_reininghaus_schoeller_15} can indeed be generalized to the
case of $N>2$ levels with a great potential for a variety of new interesting nonequilibrium fixed point
models where the low-energy behaviour in the strong coupling regime is still unknown.  

The paper is organized as follows. In Section~\ref{sec:models} we will derive various effective models.
We will set up effective tunneling models in Section~\ref{sec:tunneling_model} and the effective model 
in the cotunneling regime in Section~\ref{sec:cotunneling_regime}. The fixed point model is obtained  
via a poor man scaling analysis in Section~\ref{sec:pms_N} for arbitrary $N$. In Section~\ref{sec:pms_quarks} 
we consider the particular case $N=3$ and will set up the relation to the representation of the $SU(3)$-group 
and the physical picture in terms of $F$-spin interactions. In Section~\ref{sec:nrg} we will use the 
NRG method to confirm the $SU(3)$-symmetric fixed point model
in the linear response regime. Finally, in Section~\ref{sec:nonequilibrium_fixed_point} we analyse 
the nonequilibrium properties of the fixed point model in the perturbative regime of large voltage via
a golden rule approach. The general formulas are derived in Section~\ref{sec:golden_rule} and the
magnetization and the current are calculated as function of characteristic dot parameters for the
case of two reservoirs in Section~\ref{sec:magnetization} where the smoking gun for the detection of
the fixed point model is derived. We close with a summary of our results in Section~\ref{sec:summary}.
We use units $e=\hbar=1$ throughout this paper.

\begin{figure}
  \centering
  \resizebox {\columnwidth} {!}{
\begin{tikzpicture}
\node[draw,thick,rectangle,fill=gray!50!white,minimum width=3.5cm,minimum height=4.0cm] at (0,0) {};
\begin{scope}[rotate=-90]
\filldraw[fill=gray!50!white,thick] (-3,-5.75) parabola bend (0,-2.75) (3,-5.75);
\end{scope}
\begin{scope}[rotate=90]
\filldraw[fill=gray!50!white,thick] (-3,-5.75) parabola bend (0,-2.75) (3,-5.75);
\end{scope}
\draw[double,->,very thick,Green] (-4.3,-0.5) -- (-3.8,1.5);
\node[draw=none,fill=none] at (-5.0,0.0) {\color{Green} \large $p_L$, $q_L$};
\node[draw=none,fill=none] at (-4.5,1.5) {\color{Green} \large $\underline{\hat{f}}_L$};
\node[draw=none,fill=none] at (-4.75,-1.5) {\color{Green} \large $\mu_L = e {V \over 2}$};
\draw[double,->,very thick,Green] (4.3,-0.5) -- (4.0,1.5);
\node[draw=none,fill=none] at (5.0,0.0) {\color{Green} \large $p_R$, $q_R$};
\node[draw=none,fill=none] at (4.7,1.5) {\color{Green} \large $\underline{\hat{f}}_R$};
\node[draw=none,fill=none] at (4.75,-1.5) {\color{Green} \large $\mu_R = - e {V \over 2}$};
\coordinate (ul) at (-0.7,1.5);
\coordinate (ud) at (-0.125,1.5);
\coordinate (ur) at (1.0,1.5);
\coordinate (ml) at (-0.7,0.8);
\coordinate (md) at (0.5,0.8);
\coordinate (mr) at (1.0,0.8);
\coordinate (dl) at (-0.7,0.1);
\coordinate (dd) at (-0.125,0.1);
\coordinate (dr) at (1.0,0.1);
\coordinate (sl) at (-0.7,-1.5);
\coordinate (sr) at (1.0,-1.5);
\coordinate (sd) at (0.5,-1.5);
\coordinate (ut) at (-0.9,1.5);
\coordinate (dt) at (-0.9,0.1);
\coordinate (rl) at (-3.5,2.3);
\coordinate (rld) at (-1.0,2.3);
\coordinate (rr) at (3.5,2.3);
\coordinate (rrd) at (1.0,2.3);
\draw[very thick,Red] (ul)-- (ur);
\draw[loosely dashed,very thick,Red] (ml)-- (mr);
\draw[very thick,Red] (dl)-- (dr);
\draw[very thick,Red] (sl)-- (sr);
\draw[<->,very thick,Red] (md)-- (sd);
\draw[<->,very thick,Red] (ud)-- (dd);
\draw[<->,very thick,Red] (ut) to [bend right] (dt);
\draw[<->,very thick,Blue] (rl) to [bend left] (rld);
\draw[<->,very thick,Blue] (rrd) to [bend left] (rr);
\node[draw=none,fill=none] at (1.35,1.5) {\color{Red} \large $\ket{u}$};
\node[draw=none,fill=none] at (1.35,0.1) {\color{Red} \large $\ket{d}$};
\node[draw=none,fill=none] at (1.35,-1.5) {\color{Red} \large $\ket{s}$};
\node[draw=none,fill=none] at (0.25,1.1) {\color{Red} \large $h_z$};
\node[draw=none,fill=none] at (-1.4,0.8) {\color{Red} \large $\vec{h}_\perp$};
\node[draw=none,fill=none] at (0.825,-0.75) {\color{Red} \large $\Delta$};
\node[draw=none,fill=none] at (-2.25,3.0) {\color{Blue} \large $\Gamma_L=x_L \Gamma$};
\node[draw=none,fill=none] at (2.25,3.0) {\color{Blue} \large $\Gamma_R=x_R \Gamma$};
\end{tikzpicture}
}
  \caption{(Color online)
  Sketch of the effective model of two $F$-spin polarized leads $\alpha=L,R$ coupled to  
  a $3$-level quantum dot via flavor-conserving tunneling rates $\Gamma_{L,R}=x_{L,R}\Gamma$. 
  $\mu_{L,R}=\pm e V/2$ denote the chemical potentials of the leads with
  eight-dimensional $F$-spins $\underline{f}_{L,R}$ characterized by 
  an isospin polarization $p_{L,R}$ and hypercharge polarization $q_{L,R}$. $\vec{h}=(\vec{h}_\perp,h_z)$
  with $\vec{h}_\perp=(h_x,h_y)$ denotes the magnetic field acting on the two isospin dot levels 
  (up and down quark), and $\Delta$ is the level spacing between the strange quark and the average
  of the two isospin levels.}
  \label{fig:model}
\end{figure}
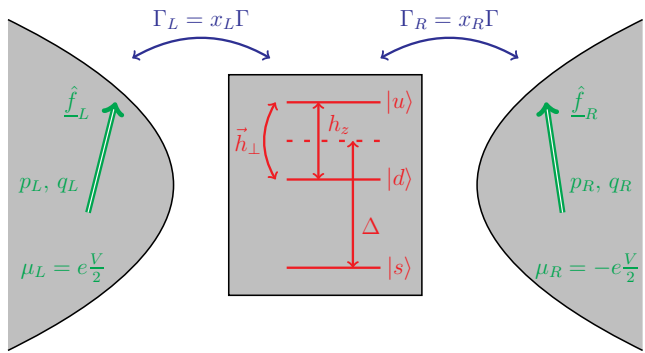

\section{Derivation of effective models}
\label{sec:models}

In this section we start from a quantum dot with $N$ levels coupled via a generic tunneling matrix to
$N_{\text{res}}$ multi-channel noninteracting reservoirs in grandcanonical
equilibrium. We show in Subsection~\ref{sec:tunneling_model} that this model is equivalent 
to an effective one where the number of channels in each reservoir is the same as the number $N$ 
of the quantum dot levels. For the special case of $N=3$ this sets the basis to use a notation in 
terms of three flavor states for the three channels and to characterize the reservoirs by 
rotated $F$-spins with a certain isospin and hypercharge polarization. In addition we will set up 
various effective tunneling models and characterize the properties of the central fixed point model
derived in Sections~\ref{sec:cotunneling_regime}-\ref{sec:pms_quarks} for the cotunneling regime, 
where the number $N_\text{dot}$ of particles on the dot is fixed to
$N_\text{dot}=1$, such that only flavor fluctuations 
via virtual intermediate states can occur. In this regime we will derive an effective model 
describing flavor fluctuations and propose the fixed point model from a poor man scaling analysis.

\subsection{Effective tunneling models}
\label{sec:tunneling_model}

The starting point is a quantum dot consisting of $N$ levels characterized by some quantum number 
$l=1,2,\dots,N$, together with a Coulomb energy $E_{N_\text{dot}}$ depending only on the total 
particle number operator $N_\text{dot}=\sum_l c^\dagger_l c_l$ of the dot
\begin{align}
\label{eq:h_dot}
H_\text{dot}\,&=\,\sum_{ll'} h_{ll'} c^\dagger_l c_{l'} \,+\, E_{N_\text{dot}} \quad,\\
\label{eq:E_N}
E_{N_\text{dot}}\,&=\, E_C(N_\text{dot}-n_x)^2\quad,
\end{align}
where $c^\dagger_l/c_l$ are the creation/annihilation operators of the single-particle states of the dot.
The charging energy $E_c$ is assumed to be the 
largest energy scale in the problem such that, for small $h_l$, the parameter $n_x$ determines the 
occupation of the dot. If $n_x=n$ is integer, the ground state will be dominated 
by $N_\text{dot}=n$, whereas for half-integer $n_x=n+{1\over 2}$, states with
$N_\text{dot}=n,n+1$ are degenerate w.r.t. the Coulomb interaction. For convenience we define the gate voltage by
\begin{align}
\label{eq:gate_voltage}
V_g\,=\,E_c (2n_x - N) \quad,
\end{align}
such that $V_g=0$ (or $n_x=N/2$) defines the particle-hole symmetric point for $h_{ll'}=0$. 
With this definition we can also write the dot Hamiltonian in second quantized form as
\begin{align}
\label{eq:h_dot_second_quantization}
H_\text{dot}\,=\, \sum_{ll'} \tilde{h}_{ll'} c^\dagger_l c_{l'} \,+\,
{U\over 2} \sum_{ll'} c^\dagger_l c^\dagger_{l'} c_{l'} c_l \quad,
\end{align}
with $U=2E_c$ and $\tilde{h}_{ll'} = h_{ll'} -(V_g + (U/2)(N-1))\delta_{ll'}$. 

The quantum dot is coupled via a generic tunneling matrix to several infinitely large reservoirs 
$\alpha=1,2,\dots,N_{\text{res}}$ kept at grandcanonical equilibrium with temperature $T$ and 
chemical potential $\mu_\alpha$, such that the total Hamiltonian reads 
\begin{align}
\label{eq:H_total}
H_{\text{tot}}\,=\,H_\text{dot}\,+\,H_{\text{res}}\,+\,H_T\quad,
\end{align}
with the reservoir Hamiltonian
\begin{align}
\label{eq:H_res}
H_{\text{res}}\,=\,\sum_{\alpha\nu_\alpha k}\,\epsilon_{\alpha\nu_\alpha k}\,a^\dagger_{\alpha\nu_\alpha k} a_{\alpha\nu_\alpha k}\quad,
\end{align}
and the tunneling Hamiltonian
\begin{align}
\label{eq:H_tun}
H_T\,=\,{1\over\sqrt{\rho^{(0)}}}\,\sum_{\alpha\nu_\alpha l k}\,
\Big\{t^{\alpha}_{\nu_\alpha l} a^\dagger_{\alpha\nu_\alpha k} c_l \,+\, 
(t^{\alpha}_{\nu_\alpha l})^* c^\dagger_l a_{\alpha\nu_\alpha k} \Big\}
\quad.
\end{align}
Here, $\nu_\alpha=1,2,\dots,N_\alpha$ is the channel index for reservoir $\alpha$ (with $N_\alpha$ channels 
in total), $\epsilon_{\alpha\nu_\alpha k}$ is the band dispersion of reservoir $\alpha$ for 
channel $\nu_\alpha$ relative to the chemical potential $\mu_\alpha$ and labelled by $k$ 
(which becomes continuous in the thermodynamic limit), and 
$t^{\alpha}_{\nu_\alpha l}$ is the tunneling matrix between the dot and reservoir $\alpha$. 
$\rho^{(0)}$ is some average density of states (d.o.s.) in the reservoirs,
which is set to $\rho^{(0)}=1$ in the following
defining the energy units. In vector-matrix notation, the tunneling Hamiltonian can be written in a more
compact form as
\begin{align}
\label{eq:H_T_compact}
H_T\,=\,\sum_{\alpha k}\Big\{\uline{a}^\dagger_{\alpha k} \,\dunderline{t}_\alpha\,\uline{c} \,+\,
\uline{c}^\dagger \,\dunderline{t^\dagger}_\alpha\,\uline{a}_{\alpha k}\Big\}\quad,
\end{align} 
where $\uline{c}^\dagger=(c^\dagger_1,\dots,c^\dagger_{N})$, 
$\uline{a}^\dagger_{\alpha k}=(a^\dagger_{\alpha 1 k},\dots,a^\dagger_{\alpha N_\alpha k})$, and 
$\dunderline{t}_\alpha$ is a $N_\alpha\times N$-matrix with matrix elements
$t^{\alpha}_{\nu_\alpha l}$. For convenience we have taken here a tunneling matrix independent of $k$ which is
usually a very good approximation for rather flat reservoir bands on the scale of the low energy scales of
interest. 

Using Keldysh formalism it is straightforward \cite{caroli_etal_72,meir_wingreen_92} to relate
the stationary current $I_{\alpha\nu_\alpha}$ in reservoir $\alpha$ and channel $\nu_\alpha$ to the 
stationary nonequilibrium greater/lesser Green's functions $G^{\gtrless}_{ll'}(\omega)$ of the dot via
\begin{align}
\nonumber
I_{\alpha\nu_\alpha}\,&=\,{e\over h}\int d\omega \,\text{Tr} \,\dunderline{\Gamma}_{\alpha\nu_\alpha}
\Big\{(1-f_\alpha(\omega)) \, i \dunderline{G}^<(\omega)\,+\\
\label{eq:current}
&\hspace{3cm}
+\,f_\alpha(\omega) \,i \dunderline{G}^>(\omega)\Big\}\quad,
\end{align}
where $\text{Tr}$ denotes the trace over the single-particle states of the dot, 
$f_\alpha(\omega)=(e^{\beta(\omega-\mu_\alpha)}+1)^{-1}$ is the Fermi function of reservoir $\alpha$, and
the $N\times N$-hybridization matrix $\dunderline{\Gamma}_{\alpha\nu_\alpha}$ is defined by
\begin{align}
\label{eq:hybridization_matrix_channel_resolved}
(\dunderline{\Gamma}_{\alpha\nu_\alpha})_{ll'}\,=\,
2\pi \rho_{\alpha\nu_\alpha} (t_{\nu_\alpha l}^\alpha)^* t_{\nu_\alpha l'}^\alpha \quad.
\end{align}
Here, $\rho_{\alpha\nu_\alpha}=\sum_k \delta(\omega-\epsilon_{\alpha\nu_\alpha k})$ denotes the d.o.s. in 
reservoir $\alpha$ for channel $\nu_\alpha$, which is assumed to be rather flat so that the energy 
dependence can be neglected. The influence of the reservoirs and the tunneling on the Green's functions
is determined by the reservoir part of the lesser/greater self-energy given by
\begin{align}
\label{eq:self_energy_lesser}
\dunderline{\Sigma}^<_{\text{res}}(\omega)\,&=\,i \sum_\alpha f_\alpha(\omega)\dunderline{\Gamma}_{\alpha}\quad,\\
\label{eq:self_energy_greater}
\dunderline{\Sigma}^>_{\text{res}}(\omega)\,&=\,-i \sum_\alpha (1-f_\alpha(\omega))\dunderline{\Gamma}_{\alpha}\quad,
\end{align}
where
\begin{align}
\label{eq:hybridization_function}
\dunderline{\Gamma}_\alpha\,=\,\sum_{\nu_\alpha} \dunderline{\Gamma}_{\alpha\nu_\alpha}
\,=\,2\pi \,\dunderline{t}_\alpha^\dagger \,\dunderline{\rho}_\alpha \,\dunderline{t}_\alpha 
\end{align}
is the hybridization matrix for reservoir $\alpha$ including all channels and 
$(\dunderline{\rho}_\alpha)_{\nu_\alpha\nu^\prime_\alpha}=\delta_{\nu_\alpha\nu^\prime_\alpha}\rho_{\alpha\nu_\alpha}$ is the
diagonal matrix for the d.o.s. of reservoir $\alpha$. As a consequence we see that 
the Green's functions depend on the reservoirs and the tunneling matrix only via the
hybridization matrices $\dunderline{\Gamma}_\alpha$ of all the reservoirs. Thus, two models with
the same hybridization matrices give exactly the same Green's functions. Once the Green's functions
are known, the channel-resolved currents $I_{\alpha\nu_\alpha}$ can be calculated from (\ref{eq:current}), where
the channel-resolved hybridization matrix $\dunderline{\Gamma}_{\alpha\nu_\alpha}$ of the concrete model
under consideration has to be inserted. The stationary expectation values of 
single-particle operators of the dot can be directly calculated from the lesser Green's functions via 
$\langle c_{l'}^\dagger c_l \rangle = {1\over 2\pi i}\int d\omega G^<_{ll'}(\omega)$ and thus are exactly 
the same for two models with the same hybridization matrices
$\dunderline{\Gamma}_\alpha$. 

We note that for the equilibrium case, where all Fermi functions of the reservoirs are the same, the reservoir 
self-energies involve only the total hybridization matrix
\begin{align}
\label{eq:hybridization_matrix_total}
\dunderline{\Gamma}\,=\,\sum_\alpha \,\dunderline{\Gamma}_\alpha\quad,
\end{align}
with the result that the equilibrium Green's functions are the same for two models with the same 
$\dunderline{\Gamma}$. However, the current in linear response can not be related to the single-particle 
Green's functions in equilibrium via (\ref{eq:current}) since also the Green's functions have to be 
expanded in the voltages. A special case is the one of proportional couplings where it is assumed that
$\dunderline{\Gamma}_\alpha = x_\alpha \dunderline{\Gamma}$ with $\sum_\alpha x_\alpha=1$.
Using current conservation $\sum_\alpha I_\alpha=0$, with $I_\alpha=\sum_{\nu_\alpha} I_{\alpha\nu_\alpha}$ denoting
the total current in reservoir $\alpha$, we get in this case from (\ref{eq:current}) the Landauer-B\"uttiker
type formula \cite{meir_wingreen_92}
\begin{align}
\label{eq:proportional_coupling}
I_\alpha\,=\,{e\over h}\sum_{\beta\ne\alpha}\int d\omega \,T_{\alpha\beta}(\omega) (f_\alpha-f_\beta)(\omega)\quad,
\end{align}
with the transmission probability
\begin{align}
\label{eq:transmission}
T_{\alpha\beta}(\omega)\,=\,2\pi \,x_\alpha x_\beta 
\text{Tr} \,\dunderline{\Gamma}\,\, \dunderline{\rho}(\omega)\quad,
\end{align}
where $\dunderline{\rho}(\omega)={i\over 2\pi}(\dunderline{G}^R-\dunderline{G}^A)(\omega)$ is the
spectral density on the dot. From this formula one can see that in linear response, where 
$(f_\alpha-f_\beta)(\omega)\approx -f'(\omega)(\mu_\alpha-\mu_\beta)$, one needs only the spectral density 
in equilibrium and, with $\mu_\alpha=-eV_\alpha$, the current can be written as
\begin{align}
\label{eq:current_linear respons}
I_\alpha\,=\,\sum_\beta G_{\alpha\beta} (V_\beta-V_\alpha)\quad,
\end{align}
with the conductance tensor
\begin{align}
\label{eq:conductance_tensor}
G_{\alpha\beta}\,=\,-{e^2\over h}\,\int d\omega\,T_{\alpha\beta}(\omega) \, f'(\omega)\quad.
\end{align}

With the knowledge that the hybridization matrices $\dunderline{\Gamma}_\alpha$ are the only input we need
to characterize the reservoirs and the tunneling matrix, we can now proceed to define effective models with
the same hybridization matrices. Since $\dunderline{\Gamma}_\alpha$ is a positive definite hermitian matrix, 
we can diagonalize it with a unitary matrix $\dunderline{U}_\alpha$ 
\begin{align}
\label{eq:gamma_diagonalize}
\dunderline{\Gamma}_\alpha\,=\,\dunderline{U}_\alpha\,\dunderline{\Gamma}_\alpha^\text{d}\,
\dunderline{U}_\alpha^\dagger 
\end{align}
where $(\dunderline{\Gamma}_\alpha^\text{d})_{ll'}=\delta_{ll'}\Gamma_{\alpha l}$ is a diagonal matrix with 
positive eigenvalues $\Gamma_{\alpha l}=2\pi t_{\alpha l}^2 > 0$. We exclude here the exotic case that one 
of the eigenvalues $\Gamma_{\alpha l}$ is zero since this would mean that one of the reservoir channels 
effectively decouples from the system. Following Ref.~\onlinecite{goettel_reininghaus_schoeller_15}, we can 
write the hybridization matrix in two equivalent forms by shifting the whole information either to an
effective tunneling matrix or to an effective d.o.s. of the reservoirs. In the first case we 
introduce an effective tunneling matrix $\dunderline{t}_\alpha^\text{eff}$ by
\begin{align}
\label{eq:tunneling_effective}
(\dunderline{t}_\alpha^\text{eff})_{ll'}\,=\,t_{\alpha l} (\dunderline{U}^\dagger_\alpha)_{ll'}\quad,
\end{align}
and get
\begin{align}
\label{eq:gamma_effective_tunneling}
\dunderline{\Gamma}_\alpha\,=\,
2\pi\,(\dunderline{t}_\alpha^\text{eff})^\dagger\,\dunderline{t}_\alpha^\text{eff}\quad.
\end{align}
Since $\dunderline{t}_\alpha^\text{eff}$ is a $N\times N$-matrix this effective model consists of 
reservoirs which have exactly the same number $N$ of channels as we have levels on the dot, i.e. 
the quantum number on the dot is also the quantum number labelling the channels in the effective
reservoirs but this quantum number is in general not conserved by tunneling. Comparing
(\ref{eq:gamma_effective_tunneling}) to (\ref{eq:hybridization_function}), we see that the
effective d.o.s. in the reservoirs is unity, i.e. we consider unpolarized reservoirs. 

In the second case, we define an effective d.o.s. $\dunderline{\rho}^{\text{eff}}_\alpha$ 
in reservoir $\alpha$ by
\begin{align}
\label{eq:effective_dos}
\dunderline{\rho}^{\text{eff}}_\alpha\,=\,
N\,\dunderline{U}_\alpha (\dunderline{\Gamma}^d_\alpha/\Gamma_\alpha)
\dunderline{U}_\alpha^\dagger\quad,
\end{align}
with $\Gamma_\alpha=\sum_l \Gamma_{\alpha l}$. Defining an average tunneling matrix element $t_\alpha>0$ by
$t^2_\alpha={1\over N}\sum_l t_{\alpha l}^2$, we can then write the hybridization matrix as
\begin{align}
\label{eq:gamma_effective_dos}
\dunderline{\Gamma}_\alpha\,=\,2\pi t_\alpha^2 \,\,\dunderline{\rho}^{\text{eff}}_\alpha\quad.
\end{align}
In this case the effective tunneling matrix is proportional to unity, the tunneling 
conserves the flavor and is flavor-independent. In contrast, the effective d.o.s. 
contains the whole nontrivial information of the hybridization matrix and describes a 
unitary transformation of the diagonal matrix 
$N \, \dunderline{\Gamma}^{\text{d}}_\alpha/\Gamma_\alpha$. The latter matrix can be decomposed in a 
basis of all diagonal matrices and the coefficients can be interpreted as physical parameters
characterizing the effective reservoirs. Using 
$\text{Tr}\,\dunderline{\Gamma}^\text{d}_\alpha = \Gamma_\alpha$ we get for $N=2$ 
\begin{align}
\label{eq:rho_decomp_N=2}
2 \, \dunderline{\Gamma}^d_\alpha/\Gamma_\alpha \,=\, \dunderline{\mathbbm{1}}_2 + p_\alpha \dunderline{\sigma}_z\quad,
\end{align}
where $\sigma_z$ is the Pauli matrix in $z$-direction and $p_\alpha$ describes the spin polarization
in reservoir $\alpha$. Since the matrix has only positive diagonal elements we get the 
condition $-1< p_\alpha <1$. If one orders the eigenvalues according to 
$\Gamma_{\alpha 1}\ge\Gamma_{\alpha 2}$ one gets $0<p_\alpha<1$. 

For $N=3$ we obtain
\begin{align}
\label{eq:rho_decomp_N=3}
3 \, \dunderline{\Gamma}^\text{d}_\alpha/\Gamma_\alpha \,=\, \dunderline{\mathbbm{1}}_3 + p_\alpha \dunderline{\lambda}_3
+ {q_\alpha\over\sqrt{3}}\dunderline{\lambda}_8 \quad,
\end{align}
where
\begin{align}
\label{eq:lambda_3}
\dunderline{\lambda}_3\,&=\,
\left(\begin{array}{ccc} 1  & 0 & 0 \\ 0 & -1 & 0 \\ 0 & 0 & 0 \end{array}\right)\quad,\\
\label{eq:lambda_8}
\dunderline{\lambda}_8\,&=\,{1\over \sqrt{3}}\left(\begin{array}{ccc} 
1  & 0 & 0 \\ 0 & 1 & 0 \\ 0 & 0 & -2 \end{array}\right)\quad,
\end{align}
are the two diagonal generators of the $SU(3)$ group, 
describing the isospin in $z$-direction of the up/down quark and the hypercharge operator 
$\dunderline{Y}={1\over\sqrt{3}}\dunderline{\lambda}_8$, respectively. Therefore we
interpret $p_\alpha$ as the isospin polarization and $q_\alpha$ as the hypercharge polarization characterizing the
reservoirs in the $3$-channel case. The fact that all matrix elements of (\ref{eq:rho_decomp_N=3}) are
positive leads to the two conditions
\begin{align}
\label{eq:p_q_condition}
|p_\alpha|\,<\, 1 + {q_\alpha \over 3}\quad,\quad
0\,<\,1+{q_\alpha \over 3}\,<\,{3\over 2} \quad.
\end{align}
If one orders the eigenvalues according to 
$\Gamma_{\alpha 1}\ge\Gamma_{\alpha 2}\ge\Gamma_{\alpha 3}$ one gets $0<p_\alpha<q_\alpha<3/2$.
 
The unitary transformation $\dunderline{U}_\alpha$ describes a
rotation of the direction of the spin-${1\over 2}$ in the $N=2$ case, and a rotation of the 
eight-dimensional $F$-spin with $\dunderline{F}_i={1\over 2}\dunderline{\lambda}_i$ 
for $N=3$, see Fig.~\ref{fig:model} for an illustration. 
Thus, the form (\ref{eq:gamma_effective_dos}) allows for a nice physical interpretation in
terms of physical parameters characterizing the reservoirs. For
$N=3$, we can label the three flavors of the reservoirs and the dot by $l=u,d,s$
for the up, down and strange quark and describe with the form (\ref{eq:gamma_effective_dos}) a system
where the flavor is conserved in tunneling with equal tunneling amplitudes for all flavors. 
However, the polarization $p_\alpha$ of the isospin described by the up and down quark and the
hypercharge polarization $q_\alpha$ can be different for each reservoir, and the $F$-spins in the reservoirs can all 
be rotated relative to the $F$-spin of the dot. This naturally generalizes the effective spin-valve model
set up in Ref.~\onlinecite{goettel_reininghaus_schoeller_15} for $N=2$ to the $N=3$ case, which is the
main subject of this paper. 

The form (\ref{eq:gamma_effective_tunneling}) in terms of an effective tunneling matrix allows for
another representation of the hybridization matrix which will turn out to be crucial to interpret
the fixed point model derived in Section~\ref{sec:pms_N} for the cotunneling regime. 
Taking all effective tunneling matrices together in a $N\cdot N_\text{res}\times N$-matrix
\begin{align}
\label{eq:total_tunneling}
\dunderline{t}^\text{eff}\,=\,
\left(\begin{array}{c} \dunderline{t}_1^\text{eff} \\ \cdot \\ \cdot \\ \cdot \\
\dunderline{t}_{N_\text{res}}^\text{eff} \end{array}\right)\quad,
\end{align}
we can write this matrix via a singular value decomposition as
\begin{align}
\label{eq:svd}
\dunderline{t}^\text{eff}\,=\,\dunderline{V}\,
\left(\begin{array}{c} \dunderline{\gamma} \\ 0 \end{array}\right)
\dunderline{W}^\dagger\quad,
\end{align}
where $\dunderline{V}$ is a unitary $N \cdot N_\text{res} \times N\cdot N_\text{res}$-matrix, 
$\dunderline{\gamma}$ is a $N\times N$-diagonal matrix containing the positive singular values 
$\gamma_1 \ge \gamma_2 \ge \dots \ge \gamma_N > 0$, and $\dunderline{W}$ is a 
unitary $N\times N$-matrix. We assume here that $N$ singular values exist, excluding exotic 
cases where some channels decouple effectively from the system. As a consequence, we can express
all effective tunneling matrices in terms of the singular value matrix $\dunderline{\gamma}$ 
as follows 
\begin{align}
\label{eq:t_gamma}
\dunderline{t}_\alpha^\text{eff}\,=\,\dunderline{V}_\alpha \,\dunderline{\gamma} 
\,\, \dunderline{W}^\dagger \quad,
\end{align}
where $\dunderline{V}_\alpha$ are the $N\times N$-matrices occuring in the first $N$ columns of 
$\dunderline{V}$, which are in general {\it not} unitary. However,
since $\dunderline{V}$ is unitary, we note the important property
\begin{align}
\label{eq:V_property}
\sum_\alpha\,\dunderline{V}_\alpha^\dagger\,\dunderline{V}_\alpha\,=\,\dunderline{\mathbbm{1}} \quad.
\end{align}
The unitary matrix $\dunderline{W}$ can be eliminated by transforming the basis of the
single-partice states of the dot using new field operators 
$\uline{c}'=\dunderline{W}^\dagger\uline{c}$, such that the dot Hamiltonian 
(\ref{eq:h_dot}) and the tunneling Hamiltonian (\ref{eq:H_T_compact}) obtain the form
\begin{align}
\label{eq:H_prime}
H_\text{dot}\,&=\,(\uline{c}^\prime)^\dagger \dunderline{h}^\prime \uline{c}^\prime 
\,+\, E_{N_\text{dot}} \quad,\\
\label{eq:H_T_compact_new}
H_T\,&=\,\sum_{\alpha k}
\Big\{\uline{a}^\dagger_{\alpha k} \,(\dunderline{t}^\text{eff}_\alpha)^\prime\,\uline{c}' \,+\,
(\uline{c}^\prime)^\dagger \,((\dunderline{t}^\text{eff}_\alpha)^\prime)^\dagger\,\uline{a}_{\alpha k}\Big\}\quad,
\end{align}
with $\dunderline{h}'=\dunderline{W}^\dagger\,\dunderline{h}\,\dunderline{W}$ and
\begin{align}
\label{eq:t_lambda_no_W}
(\dunderline{t}_\alpha^\text{eff})^\prime\,=\,\dunderline{V}_\alpha \,\dunderline{\gamma} \quad.
\end{align}
For simplicity we will drop the prime in the following and replace $\dunderline{h}'\rightarrow\dunderline{h}$
and $(\dunderline{t}_\alpha^\text{eff})^\prime\rightarrow \dunderline{t}_\alpha^\text{eff}$ keeping in 
mind that these matrices result from the matrices of the original model by transforming the dot channels
with the unitary matrix $\dunderline{W}$.

In terms of the effective tunneling matrices (\ref{eq:t_lambda_no_W}) the
hybridization matrices (\ref{eq:gamma_effective_tunneling}) obtain the form
\begin{align}
\label{eq:gamma_svd}
\dunderline{\Gamma}_\alpha\,=\, 2\pi\,\dunderline{\gamma}\,\,
\dunderline{V}_\alpha^\dagger \,\, \dunderline{V}_\alpha \,\,\dunderline{\gamma} \quad.
\end{align}
This form is of particular interest since it separates the hybridization matrix in a part
$\dunderline{\gamma}$ which is independent of the reservoirs and a reservoir-dependent part 
$\dunderline{V}_\alpha^\dagger \,\dunderline{V}_\alpha$. Comparing (\ref{eq:gamma_svd}) with 
(\ref{eq:hybridization_function}) we can interpret $\dunderline{\gamma}$ as an effective 
tunneling matrix  which conserves the flavor index and is the same for all reservoirs. This 
effective tunneling matrix contains the information of the eigenvalues $\Gamma_l=2\pi\gamma_l^2$ 
of the total hybridization matrix since we get from (\ref{eq:V_property}) 
\begin{align}
\label{eq:gamma_total_diagonal}
\dunderline{\Gamma}\,=\,\sum_\alpha\,\dunderline{\Gamma}_\alpha\,=\,
2\pi \,\, \dunderline{\gamma}^2\quad.
\end{align}
The reservoir-dependent part $\dunderline{V}_\alpha^\dagger \,\dunderline{V}_\alpha$ can be interpreted 
as an effective d.o.s. of the reservoirs. Taking $N=3$ and decomposing this hermitian matrix in the 
basis of the $F$-spin generators $\dunderline{F}_i={1\over 2}\dunderline{\lambda}_i$ of $SU(3)$ we get
\begin{align}
\label{eq:decompose_rho}
\dunderline{V}^\dagger_\alpha \, \dunderline{V}_\alpha\,=\,
x_\alpha\,\Big(\dunderline{\mathbbm{1}}\,+\,\sum_{i=1}^8\,d_\alpha^i\,\dunderline{F}_i\Big) \quad,
\end{align}
with real coefficients $x_\alpha$ and $d_\alpha^i$ which, due to (\ref{eq:V_property}), fulfil the
property
\begin{align}
\label{eq:x_d_property}
\sum_\alpha x_\alpha\,=\,1 \quad,\quad \sum_\alpha x_\alpha d_\alpha^i\,=\,0 \quad.
\end{align}
This means that the sum of the $F$-spins of all reservoirs is zero. A similiar property holds for
arbitrary $N$. In equilibrium, where all chemical potentials $\mu_\alpha=\mu$ are the same and all 
reservoirs can be taken together to one big reservoir, this means that an unpolarized reservoir with
$SU(3)$-symmetry couples to the dot. However, since the effective tunneling matrix elements $\gamma_l$
still depend on the flavor index, $SU(3)$-symmetry does not hold for the total system even in 
equilibrium. 

Most importantly, we will see in Section~\ref{sec:pms_N} by a poor man scaling analysis 
in the cotunneling regime of a singly occupied dot $N_\text{dot}=1$ that a generic fixed point model 
with an isotropic matrix $\dunderline{\gamma}=\gamma\dunderline{\mathbbm{1}}$ emerges,
such that the effective tunneling matrix (\ref{eq:t_lambda_no_W}) reads
\begin{align}
\label{eq:t_fixed_point}
\dunderline{t}_\alpha^\text{eff}\,=\,\gamma \dunderline{V}_\alpha  \quad.
\end{align}
$\gamma>0$ can be related to an isotropic Kondo coupling $J$ via 
\begin{align}
\label{eq:gamma_J}
\gamma^2\,=\,{1 \over 4} J D \quad,\quad {1\over D}\,=\,{1\over 2}\left({1\over E_0}+{1\over E_2}\right)
\quad,
\end{align}
where $E_{N_\text{dot}}$ is given by (\ref{eq:E_N}) and $J$ fulfils the poor man scaling equation
\begin{align}
\label{eq:pms_J}
{d J \over d\Lambda}\,=\,-{N\over 2}{J^2\over\Lambda} \quad,
\end{align}
with $\Lambda$ denoting the effective band width. In addition, a special potential scattering term
emerges in the original tunneling model at the fixed point which is given by
\begin{align}
\label{eq:potential_scattering}
V_\text{sc}\,=\,v_\text{sc}\sum_{kk'}\sum_{\alpha\alpha'}
:\underline{a}^\dagger_{\alpha k} \,\dunderline{V}_\alpha\,\,\dunderline{V}^\dagger_{\alpha'}\,
\underline{a}_{\alpha' k'}: \quad,
\end{align}
where
\begin{align}
\label{eq:v_sc}
v_\text{sc}\,=\,{\gamma^2\over D}\left({N-2\over N} + \delta \right)\quad,\quad
\delta\,=\,{E_0-E_2 \over E_2 + E_0}\quad,
\end{align}
with $E_0$ and $E_2$ from \eqref{eq:E_N}, 
and $:\dots:$ denotes normal-ordering. This potential scattering term vanishes for $N=2$ 
and $\delta=0$ (i.e. $n_x=1$ where $E_0=E_2$) and is
such that it cancels the potential scattering term emerging in an effective model for  
the cotunneling regime, see Section~\ref{sec:cotunneling_regime}. Due to $V_\text{sc}$
the reservoir part of the self-energy of the dot is more 
complicated than (\ref{eq:self_energy_lesser}-\ref{eq:self_energy_greater}) and does not only
depend on the hybridization matrix. However, as is shown in Appendix~\ref{app:res_self_energy}, the  
effect of $V_\text{sc}$ is just that $\gamma$ is changed to an effective 
$\tilde{\gamma}$ given by
\begin{align}
\label{eq:eff_lambda}
\tilde{\gamma}\,=\,{\gamma\over \sqrt{1+\pi^2 v_\text{sc}^2}}\quad,
\end{align}
such that the self-energies (\ref{eq:self_energy_lesser}-\ref{eq:self_energy_greater}) from the reservoirs
can be written at the fixed point with effective hybridization matrices which can be either expressed via an
effective tunneling matrix analog to (\ref{eq:gamma_effective_tunneling})
\begin{align}
\label{eq:gamma_effective_tunneling_fixed_point_model}
\dunderline{\Gamma}_\alpha\,=\,
2\pi\,(\dunderline{t}_\alpha^\text{eff})^\dagger\,\dunderline{t}_\alpha^\text{eff}\quad,\quad
\dunderline{t}_\alpha^\text{eff}\,=\,\tilde{\gamma} \dunderline{V}_\alpha  \quad,
\end{align}
such that the d.o.s. of the reservoirs is unity, or via an effective d.o.s. 
analog to (\ref{eq:gamma_effective_dos})
\begin{align}
\label{eq:gamma_effective_dos_fixed_point}
\dunderline{\Gamma}_\alpha\,&=\,2\pi \tilde{\gamma}^2\,\dunderline{\rho}_\alpha^\text{eff}\quad,\quad
\dunderline{\rho}_\alpha^\text{eff}\,=\,\dunderline{V}_\alpha^\dagger\,\,\dunderline{V}_\alpha \quad,
\end{align}
with a trivial tunneling matrix given by $\tilde{\gamma}\dunderline{\mathbbm{1}}$ which is the same for all 
reservoirs and proportional to unity w.r.t the flavor indices. The particular property of the
effective d.o.s. at the fixed point is the condition
\begin{align}
\sum_\alpha\,\dunderline{\rho}_\alpha^\text{eff}\,=\,\dunderline{\mathbbm{1}}\quad,
\end{align}
following from (\ref{eq:V_property}). This means that in contrast to the general case depicted in 
Fig.~\ref{fig:model} for $N=3$, the particular property of the fixed point model is that the sum 
over all reservoir $F$-spins is equal to zero {\it and} the tunneling matrix 
$\dunderline{\gamma}=\gamma\dunderline{\mathbbm{1}}$ is isotropic. 
As a consequence we get overall $SU(3)$-symmetry in equilibrium,
whereas in nonequilibrium the fixed point model is essentially {\it not} $SU(3)$-symmetric since the 
$F$-spins of the reservoirs are nonzero.
A similiar statement holds for any number $N$ of dot levels, generalizing the picture found in
Ref.~\onlinecite{goettel_reininghaus_schoeller_15} for $N=2$ to a generic multi-level quantum dot. 

We note that for the particular case of two reservoirs $N_\text{res}=2$ with $\alpha=L,R$, 
we get from (\ref{eq:V_property}) that 
$\dunderline{V}_L^\dagger\,\dunderline{V}_L = \dunderline{\mathbbm{1}} - \dunderline{V}_R^\dagger\,\dunderline{V}_R$,
such that we can find a common unitary matrix $\dunderline{U}_V$ which diagonalizes both
$\dunderline{V}_\alpha^\dagger\,\dunderline{V}_\alpha$ for $\alpha=L,R$
\begin{align}
\label{eq:diagonalization_VV}
\dunderline{V}_\alpha^\dagger\,\dunderline{V}_\alpha\,=\,
\dunderline{U}_V\,\,\dunderline{A}_\alpha^\text{d}\,\,\dunderline{U}^\dagger_V\quad,
\end{align}
where $\dunderline{A}_\alpha^\text{d}$ are diagonal matrices with the property
\begin{align}
\label{eq:A_property}
\sum_{\alpha=L,R}\,\dunderline{A}_\alpha^\text{d}\,=\,\dunderline{\mathbbm{1}}\quad.
\end{align}
For $N=3$, the matrix $\dunderline{A}_\alpha^\text{d}$ can be decomposed analog to (\ref{eq:rho_decomp_N=3}) as
\begin{align}
\label{eq:A_decomp_N=3}
\dunderline{A}^\text{d}_\alpha \,=\, x_\alpha \Big(\dunderline{\mathbbm{1}}_3 + p_\alpha \dunderline{\lambda}_3
+ {q_\alpha\over\sqrt{3}}\dunderline{\lambda}_8 \Big)\quad,
\end{align}
where, due to the property (\ref{eq:A_property}), we get
\begin{align}
\label{eq:x_property_two_reservoirs}
1\,&=\,x_L\,+\,x_R \quad,\\
\label{eq:p_property_two_reservoirs}
0\,&=\,x_L p_L\,+\,x_R p_R \quad,\\
\label{eq:q_property_two_reservoirs}
0\,&=\,x_L q_L\,+\,x_R q_R \quad,
\end{align}
together with $0<x_\alpha<1$ and (\ref{eq:p_q_condition}).
Thus, the hybridization matrices at the fixed point obtain the following form for two reservoirs
\begin{align}
\label{eq:gamma_fixed_point_two_reservoirs}
\dunderline{\Gamma}_\alpha\,=\, 2\pi \tilde{\gamma}^2\,\,
\dunderline{U}_V \,\, \dunderline{A}_\alpha^\text{d} \,\,\dunderline{U}^\dagger_V \quad.
\end{align}
Omitting the unitary matrix $\dunderline{U}_V$ by choosing a different single-particle basis for the 
dot states and redefining the parameters $h_{ll'}$ (analog to the transformation by the unitary matrix
$\dunderline{W}$, see (\ref{eq:H_prime}-\ref{eq:H_T_compact_new})), we get finally the diagonal form
\begin{align}
\label{eq:gamma_fixed_point_two_reservoirs_diagonal}
\dunderline{\Gamma}_\alpha\,=\, 2\pi \tilde{\gamma}^2\,\,
\dunderline{A}_\alpha^\text{d} \quad,
\end{align}
which, for $N=3$, by inserting the decomposition (\ref{eq:A_decomp_N=3}), can be written as
\begin{align}
\label{eq:gamma_fixed_point_two_reservoirs_diagonal_N=3}
\dunderline{\Gamma}_\alpha\,=\, {1\over 3}\Gamma_\alpha\,\,
\Big(\dunderline{\mathbbm{1}}_3 + p_\alpha \dunderline{\lambda}_3
+ {q_\alpha\over\sqrt{3}}\dunderline{\lambda}_8\Big) \quad,
\end{align}
with $\Gamma_\alpha=2\pi\tilde{\gamma}^2 x_\alpha$. This form 
for the hybrization matrices constitutes the central generic fixed point model for $N=3$ and two 
reservoirs in the cotunneling regime of a singly occupied dot. This will be confirmed
in Section~\ref{sec:nrg} by NRG in equilibrium and analysed in Section~\ref{sec:nonequilibrium_fixed_point} by
a golden rule approach in nonequilibrium. It generalizes the spin-valve model for 
a $2$-level quantum dot with opposite spin polarizations in the two reservoirs analysed in 
Ref.~\onlinecite{goettel_reininghaus_schoeller_15} to the case of
a $3$-level quantum dot, where the isospin and hypercharge polarizations have to be opposite in
the two reservoirs. An analog fixed point model arises for an arbitrary number of dot levels, in this case
one obtains in the two reservoirs opposite parameters corresponding to the $N-1$ diagonal generators of $SU(N)$. 
Whereas in equilibrium the fixed point model is $SU(N)$-symmetric (at least if the dot 
parmeters $h_{ll'}$ are adjusted properly, see Section~\ref{sec:nrg}) and leads generically to the
$SU(N)$-Kondo effect, the nonequilibrium fixed point model is {\it not} $SU(N)$-symmetric and generically
non-Kondo physics has to be expected. This will be analysed in Section~\ref{sec:nonequilibrium_fixed_point} in the
perturbative golden rule regime of large voltage, where we will see that zero $F$-spin 
magnetization on the dot occurs only for particular values of the dot parameters $h_{ll'}$ providing a
{\it smoking gun} for the detection of the fixed point model.

\subsection{Effective model in the cotunneling regime}
\label{sec:cotunneling_regime}

The effective model in the cotunneling regime where the particle number on the dot is fixed to 
$N_\text{dot}=1$ can easily be obtained by projecting the Hamiltonian matrix on this subspace analog 
to Brillouin-Wigner perturbation theory. Taking only one virtual process into the particle number 
sectors $N_\text{dot}=0,2$ into account we get 
\begin{align}
\label{eq:H_tot_eff}
H_\text{tot}^\text{eff}\,=\,H_\text{res}\,+\,P_1 H_\text{dot} P_1 
\,-\, :P_1 H_T Q_1 {1\over H_\text{dot}} Q_1 H_T P_1:  \quad,
\end{align}
where $P_1$ projects onto the $1$-particle subspace of the dot and $Q_1=1-P_1$. We have introduced the
normal-ordering $:\dots :$ w.r.t. the reservoir field operators since we are not interested in terms
renormalizing the dot Hamiltonian leading to effective parameters $h_{ll'}$. For $H_\text{res}$
and $H_T$ we take a model with the effective tunneling matrix (\ref{eq:t_lambda_no_W}) and the unity matrix
for the effective d.o.s. in the reservoirs, as has been discussed in Section~\ref{sec:models}. 
Inserting $H_\text{dot}$ and $H_T$ from (\ref{eq:h_dot}) and
(\ref{eq:H_T_compact}) and using $E_{N_\text{dot}}$ from \eqref{eq:E_N} we get with 
$P_1 c_l^\dagger c_{l'}P_1 = |l\rangle\langle l'|$
\begin{align}
\nonumber
H_\text{tot}^\text{eff}\,&=\,H_\text{res}\,+\,\sum_{ll'} h_{ll'}|l\rangle\langle l'|\,+\,E_1\\
&\hspace{-1cm}
-\,{1\over E_2} \sum_{\alpha\alpha'}\sum_{kk'} 
:P_1 \uline{a}^\dagger_{\alpha k} \,\dunderline{t}^\text{eff}_\alpha\,
\uline{c}\,\uline{c}^\dagger \,(\dunderline{t}^\text{eff}_{\alpha'})^\dagger\,\uline{a}_{\alpha' k'} P_1:\\
\label{eq:H_tot_eff_explicit}
&\hspace{-1cm}
-\,{1\over E_0} \sum_{\alpha\alpha'}\sum_{kk'} 
:P_1 \uline{c}^\dagger \,(\dunderline{t}^\text{eff}_{\alpha'})^\dagger\,
\uline{a}_{\alpha' k'}\,\uline{a}^\dagger_{\alpha k} \,\dunderline{t}^\text{eff}_\alpha\,\uline{c}P_1:\quad.
\end{align}
Using 
\begin{align}
\nonumber 
P_1(\uline{c}\,\uline{c}^\dagger)_{ll'}P_1 \,&=\, -|l'\rangle\langle l| + \delta_{ll'}P_1 \quad,\\
:(\uline{a}_{\alpha' k'}\,\uline{a}^\dagger_{\alpha k})_{l'l}:\,&=\,-:a^\dagger_{\alpha l k } a_{\alpha'l'k'}:\quad,
\end{align}
we get after inserting (\ref{eq:t_lambda_no_W}) for the tunneling matrix and leaving out the unimportant 
constant $E_1$
\begin{align}
\label{eq:H_tot_eff_final}
H_\text{tot}^\text{eff}\,=\,H_\text{res}\,+\,\sum_{ll'} h_{ll'}|l\rangle\langle l'|
\,+\,V_\text{eff} \quad,
\end{align}
with the effective interaction
\begin{align}
\label{eq:V_eff}
V_\text{eff}\,=\,\sum_{\alpha\alpha'}\sum_{kk'}:\uline{a}^\dagger_{\alpha k} 
\,\dunderline{V}_\alpha\,\dunderline{\hat{J}}\,\dunderline{V}_{\alpha'}^\dagger\,\uline{a}_{\alpha' k'}:\quad,
\end{align}
and
\begin{align}
\label{eq:J}
\hat{J}_{ll'}\,=\,\gamma_l\gamma_{l'}\left({2\over D}|l'\rangle\langle l| 
\,-\,{1\over E_2}\delta_{ll'}\hat{\mathbbm{1}}\right)
\quad,
\end{align}
with $2/D=1/E_0+1/E_2$, see (\ref{eq:gamma_J}). We note that the hat on $\hat{J}_{ll'}$
indicates that this object is a dot operator in the $1$-particle subspace for {\it each} fixed value 
of $l$ and $l'$, i.e. $\dunderline{\hat{J}}$ represents a $N\times N$-matrix with dot operators in 
each matrix element. By using $\hat{\mathbbm{1}}=\sum_l |l\rangle\langle l|$, a straighforward calculation
leads to the decomposition
\begin{align}
\nonumber
\hat{J}_{ll'}\,&=\,\xi_{ll'} |l'\rangle\langle l| (1-\delta_{ll'})\,+ \\
\label{eq:J_splitting}
&\hspace{-1cm}
+\,\sum_{l_1\ne l}\eta_{ll_1}\Big({1\over N}\hat{\mathbbm{1}}-|l_1\rangle\langle l_1|\Big)\delta_{ll'}
\,+\,v_l \delta_{ll'}\hat{\mathbbm{1}} \quad,
\end{align}
with
\begin{align}
\label{eq:xi_eta_ll'}
\xi_{ll'}\,&=\,{2\over D}\gamma_l\gamma_{l'} \quad,\quad
\eta_{ll'}\,=\,{2\over D}\gamma_l^2\quad,\\
\label{eq:v_l}
v_l\,&=\,-{1\over D}\gamma_l^2 ({N-2\over N} + \delta) \quad,
\end{align}
and $2\delta = D/E_2-D/E_0$, see (\ref{eq:v_sc}). We note that the bare
parameters $\eta_{ll'}$ are independent of $l'$ but obtain a strong dependence on $l'$ under the 
renormalization group flow described below. The decomposition (\ref{eq:J_splitting}) exhibits the
non-diagonal matrix $|l'\rangle\langle l|$ for $l\ne l'$, all traceless diagonal matrices 
${1\over N}\hat{\mathbbm{1}}-|l_1\rangle\langle l_1|$ for $l_1\ne l$, and the unity matrix $\hat{\mathbbm{1}}$
describing the effective potential scattering. 

We note that the effective interaction (\ref{eq:V_eff}) can also be written in terms of
reservoir field operators for a single reservoir only
\begin{align}
\label{eq:V_eff_one_reservoir}
V_\text{eff}\,=\,\sum_{kk'}:\uline{\tilde{a}}^\dagger_k \,\dunderline{\hat{J}}\,\uline{\tilde{a}}_{k'}:\quad,
\end{align}
where 
\begin{align}
\label{eq:tilde_a}
\uline{\tilde{a}}_k\,=\,\sum_\alpha\,\dunderline{V}_{\alpha}^\dagger\,\uline{a}_{\alpha k}
\end{align}
fulfil commutation relations of field operators for a single effective reservoir with $N$ flavors
due to the property (\ref{eq:V_property}). However, this is 
only possible if all the reservoirs can be taken together, i.e. they must have the same temperature
and chemical potential. In nonequilibrium this is not possible. Nevertheless, for the poor man scaling
analysis described in the next section, this form of the Hamiltonian can be applied since the poor man
scaling analysis integrates out only energy scales above the temperatures and chemical potentials of
the reservoirs.

\subsection{Poor man scaling and fixed point model for N levels}
\label{sec:pms_N}

Taking the effective Hamiltonian in the cotunneling regime in the form 
(\ref{eq:H_tot_eff_final}) and (\ref{eq:V_eff_one_reservoir}), 
we now proceed to find an effective low-energy theory by
integrating out all energy scales from the high-energy cutoff $\Lambda=D$ down to some low-energy scale 
$\Lambda_c$ defined by the largest physical low energy scale in the system set by the parameters $h_{ll'}$
of the dot Hamiltonian, the temperature $T$ of the reservoirs, and the chemical potentials $\mu_\alpha$ of
the reservoirs 
\begin{align}
\label{eq:Lambda_c}
\Lambda_c\,=\,\text{max}\Big\{\{|h_{ll'}|\}_{ll'},T,\{\mu_\alpha\}_\alpha\Big\}\quad.
\end{align}
This can be achieved by a standard poor man scaling analysis leading to the RG equations
\begin{align}
\label{eq:pms}
{d \hat{J}_{ll'} \over ds} \,=\, -\sum_{l_1} 
\Big[ \hat{J}_{ll_1},\hat{J}_{l_1 l'} \Big] \quad,
\end{align}
where $[\cdot,\cdot]$ denotes the commutator and $s=\ln{D\over \Lambda}$ is the flow parameter. 
This RG equation has obviously the two invariants
$\text{Tr}\hat{J}_{ll'}$ and $\sum_l \hat{J}_{ll}$. Defining 
\begin{align}
\label{eq:eta_v_sum}
\eta_l \,&=\, \sum_{l'\ne l} \eta_{ll'} \quad,\quad 
\eta \,=\, \sum_l\eta_l \quad,\quad v\,=\,\sum_l v_l \quad,
\end{align}
we obtain from the decomposition (\ref{eq:J_splitting}) 
\begin{align}
\label{eq:invariant_trace}
\text{Tr}\hat{J}_{ll'} \,&=\, N v_l \delta_{ll'}\quad,\\
\label{eq:invariant_diagonal_sum}
\langle l|\sum_{l'}\hat{J}_{l'l'}|l\rangle \,&=\,{1\over N}\eta \,-\, \eta_l \,+\,v \quad,
\end{align}
and get the invariants 
\begin{align}
\label{eq:v_invariant}
0\,&=\,{d\over ds}v_l \quad,\\
\label{eq:eta_invariant}
0\,&=\,{d\over ds} \Big(\eta_l\,-\,{1\over N}\eta\Big) \quad.
\end{align}
The first equation means that there is no renormalization for the potential scattering. The
second equation holds for all $l=1,\dots,N$ and gives $N-1$ independent invariants. 

Inserting the decomposition (\ref{eq:J_splitting}) in (\ref{eq:pms}) we find
after some straightforward algebra the RG equations for the parameters $\xi_{ll'}$ and  
$\eta_{ll'}$ characterizing the effective operator-valued matrix $\dunderline{\hat{J}}$
at scale $\Lambda$ in terms of (\ref{eq:J_splitting}) ($l\ne l'$ in all following equations)
\begin{align}
\label{eq:rg_gamma}
{d\xi_{ll'}\over d s}\,&=\,2\xi_{ll'}\bar{\eta}_{ll'}\,+\,
\sum_{l_1\ne l,l'}\xi_{ll_1}\xi_{l_1 l'} \quad,\\
\label{eq:rg_eta}
{d\eta_{ll'}\over d s}\,&=\,2\xi_{ll'}\xi_{l'l}\,+\,
\sum_{l_1\ne l,l'}\xi_{ll_1}\xi_{l_1 l} \quad,
\end{align}
where we defined the symmetric matrix
\begin{align}
\label{eq:bar_eta}
\bar{\eta}_{ll'}\,=\,{1\over 2}(\eta_{ll'}+\eta_{l'l}) \quad,
\end{align}
which fulfils the RG equation
\begin{align}
\label{eq:rg_eta_symmetric}
{d\bar{\eta}_{ll'}\over d s}\,&=\,2\xi_{ll'}^2\,+\,
{1\over 2}\sum_{l_1\ne l,l'}\Big(\xi_{ll_1}^2 + \xi_{l_1 l'}^2\Big) \quad,
\end{align}
since $\xi_{ll'}$ stays symmetric during the whole RG flow
\begin{align}
\label{eq:xi_symmetry}
\xi_{ll'}\,=\,\xi_{l'l} \quad.
\end{align}
These differential equations have to be solved starting from the initial conditions at $s=0$ 
given by (\ref{eq:xi_eta_ll'}). 

The RG equation for $\eta_{ll'}$ can be solved by the ansatz
\begin{align}
\label{eq:ansatz_eta}
\eta_{ll'}\,=\,\bar{\eta}_{ll'} + r_l - r_{l'}\quad,
\end{align}
where the $r_l$ are determined from the RG equations
\begin{align}
\label{eq:RG_r}
{d r_l \over ds}\,=\,{1\over 2}\sum_{l'\ne l} \xi_{ll'}^2 \quad,
\end{align}
with initial condition $r_l=\gamma_l^2/D$. Using the form (\ref{eq:ansatz_eta}) we can express the 
$N-1$ independent invariants (\ref{eq:eta_invariant}) as
\begin{align}
\label{eq:r_l_invariant}
0\,=\,{d\over ds} \Big(r\,-\,N r_l \,+\,\bar{\eta}_l \,-\, {1\over N}\bar{\eta}  \Big)\quad,
\end{align}
where we have defined in analogy to (\ref{eq:eta_v_sum}) 
\begin{align}
\label{eq:bar_eta_r_sum}
\bar{\eta}_l \,=\, \sum_{l'\ne l} \bar{\eta}_{ll'} \,\,\,,\,\,\,
\bar{\eta} \,=\, \eta \,=\, \sum_l\bar{\eta}_l \,\,\,,\,\,\, r\,=\,\sum_l r_l \quad.
\end{align}
With these invariants all $N-1$ differences $r_l-r_{l'}$ can be expressed via the
symmetric matrix $\bar{\eta}_{ll'}$ and it is only necessary to consider the RG equations 
(\ref{eq:rg_gamma}) and (\ref{eq:rg_eta_symmetric}) for the symmetric matrices 
$\xi_{ll'}$ and $\bar{\eta}_{ll'}$. As we will see in Section~\ref{sec:pms_quarks}, these 
coupling constants can be interpreted as the transverse and longitudinal Kondo couplings
$J_\perp$ and $J_z$ corresponding to the $SU(2)$-subgroup formed by the level pair $(l,l')$. 

As one can see from (\ref{eq:rg_eta}) the parameters $\eta_{ll'}$ obtain a significant
dependence on $l'$ not present in the initial condition. Furthermore, all parameters $\xi_{ll'}$ and
$\eta_{ll'}$ stay positive and increase monotonously under the RG flow until they diverge at a 
certain low-energy scale $T_K$. The fixed point is the one where all parameters are the same and 
proportional to an isotropic Kondo-like coupling $J$
\begin{align}
\label{eq:xi_eta_fixed_point}
\xi_{ll'}\,=\,\eta_{ll'}\,=\,{1\over 2} J \quad,
\end{align}
where $J$ fulfils the RG equation (\ref{eq:pms_J}) 
\begin{align}
\label{eq:J(s)_rg}
{d J\over ds}\,=\,{N\over 2}J^2 \quad\Rightarrow\quad T_K\,=\,\Lambda e^{-{2\over N J}}\,=\,\text{const}\quad.
\end{align}
$T_K$ is the energy scale where all coupling constants diverge and is called the Kondo temperature 
in the following. This scale is exponentially sensitive to the choice of the initial conditions. 
Therefore, one defines a typical initial coupling $J_0$ via
\begin{align}
\label{eq:J_0}
{4 \gamma_l^2 \over D} \,=\, y_l\,J_0\quad,\quad \sum_l y_l\,=\,1 \quad,
\end{align}
such that $y_l\sim O(1)$ are fixed parameters, and defines formally the scaling limit by 
\begin{align}
\label{eq:scaling_limit}
J_0\,\rightarrow\,0 \quad,\quad D\,\rightarrow\,\infty \quad,\quad 
T_K\,=\,\text{const} \quad.
\end{align}

Close to the fixed point we can neglect the small
potential scattering term and get from (\ref{eq:J_splitting}) the form
\begin{align}
\nonumber
\hat{J}_{ll'}\,&=\,{1\over 2}J |l'\rangle\langle l| (1-\delta_{ll'})\,+ \\
\label{eq:J_splitting_fixed_point}
&\hspace{-1cm}
+\,{1\over 2} J \sum_{l_1\ne l}\Big({1\over N}\hat{\mathbbm{1}}-|l_1\rangle\langle l_1|\Big)\delta_{ll'}
\quad,
\end{align}
which can also be written in the more compact form
\begin{align}
\label{eq:J_fixed_point_compact}
\hat{J}_{ll'}\,&=\,{1\over 2}J |l'\rangle\langle l| \,-\, 
{1 \over 2N} J \hat{\mathbbm{1}}\delta_{ll'}\quad.
\end{align}
Using this form in the effective interaction (\ref{eq:V_eff_one_reservoir}) we get at the
fixed point in the $1$-particle subspace of the dot 
\begin{align}
\nonumber
V_\text{eff}\,&=\,-\,{1\over 2N} J \sum_{kk'} : \uline{\tilde{a}}^\dagger_k\,\uline{\tilde{a}}_{k'}: \\
\label{eq:V_fixed_point}
&\hspace{1cm}
\,+\,{1\over 2}J \sum_{kk'}\sum_{ll'} c^\dagger_{l'} c_l \,:\tilde{a}^\dagger_{lk}\,\tilde{a}_{l'k'}: \quad. 
\end{align}
At the fixed point the effective interaction is obviously $SU(N)$-invariant under a common unitary 
transformation of the $N$ flavors of the reservoir and dot field operators. We note that this holds only 
in the case of the single reservoir described by the field operators $\tilde{a}_{lk}$ but not for the original
model in nonequilibrium where the reservoirs have different chemical potentials $\mu_\alpha$. In this
case one has to insert (\ref{eq:tilde_a}) in (\ref{eq:V_fixed_point}) and finds that the effective interaction
is {\it not} invariant under a common unitary transformation of all dot field operators $c_l$ and reservoir
field operators $a_{\alpha l k}$ due to the presence of the matrices $\dunderline{V}_\alpha$.

We finally show that the fixed point Hamiltonian corresponds to a projection of the effective tunneling model 
(\ref{eq:t_fixed_point}) together with the potential scattering term (\ref{eq:potential_scattering})
on the $N=1$ subspace of the dot. Comparing (\ref{eq:xi_eta_fixed_point}) with (\ref{eq:xi_eta_ll'}) 
we find that we get indeed a unity matrix for $\dunderline{\gamma}=\gamma\dunderline{\mathbbm{1}}$ with $\gamma$ 
given by (\ref{eq:gamma_J}). Furthermore, the potential scattering is absent in the fixed point model 
(\ref{eq:J_splitting_fixed_point}) and, therefore, we have to introduce the potential
scattering term (\ref{eq:potential_scattering}) in the effective tunneling model with a coupling constant 
$v_\text{sc}$ given by (\ref{eq:v_sc}) of opposite sign compared to (\ref{eq:v_l}) 
(where $\gamma_l$ is replaced by $\gamma$) such that (\ref{eq:potential_scattering}) cancels the potential 
scattering generated by projecting the effective tunneling model on the $N=1$ subspace.

\subsection{Poor man scaling in $SU(3)$-representation}
\label{sec:pms_quarks}

For the $3$-level case $N=3$, which is the main subject of this paper, it is quite instructive
to write the Hamiltonian and the poor man scaling equations also in the representation of the generators 
of the $SU(3)$-group. This provides a nice physical picture how the reservoir and dot $F$-spins are coupled
and how the interaction can be interpreted in terms of the dot and reservoir quark flavors. 

Since each matrix element $\hat{J}_{ll'}$ is an operator in the $3$-dimensional dot space we
can decompose it in the $F$-spin components $\hat{F}_i={1\over 2}\hat{\lambda}_i$ of the dot as
\begin{align}
\label{eq:generators_dot_space}
\hat{J}_{ll'}\,=\,\sum_{i=1}^8 J^i_{ll'}\hat{F}_i \,+\,v_l\delta_{ll'}\hat{\mathbbm{1}} \quad,
\end{align}
where the last term contains the potential scattering. Furthermore 
each $3\times 3$-matrix $\dunderline{J}^i$ can again be decomposed in the 
generators $\dunderline{\lambda}_j$ in reservoir space (note that we still consider here only one effective
reservoir due to the form (\ref{eq:V_eff_one_reservoir}) of the effective interaction in the poor man
scaling regime). Comparing (\ref{eq:generators_dot_space}) with (\ref{eq:J_splitting}) we find after some
straightforward algebra
\begin{align}
\label{eq:decomposition_J_124567}
\dunderline{J}^i\,&=\, J_i\,\dunderline{\lambda}_i \quad,\quad \text{for}\,\,i=1,2,4,5,6,7 \quad,\\
\label{eq:decomposition_J_3}
\dunderline{J}^3\,&=\, J_3 \,\dunderline{\lambda}_3 \,+\,J_{38}\,\dunderline{\lambda}_8
 \,+\, {2\over 3}\,c_3\,\dunderline{\mathbbm{1}} \quad,\\
\label{eq:decomposition_J_8}
\dunderline{J}^8\,&=\, J_8 \,\dunderline{\lambda}_8 \,+\, J_{83} \,\dunderline{\lambda}_3 
\,+\, {2\over 3\sqrt{3}}\,c_8\,\dunderline{\mathbbm{1}} \quad,
\end{align}
where the various coupling constants are defined by 
\begin{align}
\label{eq:J_K_12}
J_1 &= J_2 = \xi_{12} \quad,\quad K_1 = \bar{\eta}_{12} \quad,\\
\label{eq:J_K_45}
J_4 &= J_5 = \xi_{13} \quad,\quad K_4 = \bar{\eta}_{13} \quad,\\
\label{eq:J_K_67}
J_6 &= J_7 = \xi_{23} \quad,\quad K_6 = \bar{\eta}_{23} \quad,\\
\label{eq:J_3_8}
J_3\,&=\,K_1 \quad,\quad J_8\,=\,{1\over 3}\,(2K_4 + 2K_6 - K_1)\quad,\\
\label{eq:J_38}
J_{38}\,&=\,J_{83}\,=\,{1\over\sqrt{3}}\,(K_4-K_6)\quad,
\end{align}
together with the two invariants
\begin{align}
\label{eq:c_38}
c_3\,=\,{\gamma_1^2-\gamma_2^2\over D} \quad,\quad
c_8\,=\,{\gamma_1^2+\gamma_2^2-2\gamma_3^2\over D} \quad.
\end{align}
$c_3$ and $c_8$ must be invariants since
\begin{align}
\label{eq:sum_l_invariant_SU3}
\sum_l \hat{J}_{ll}\,=\,\sum_{i=1}^8 \Big(\text{Tr} \dunderline{J}^i\Big)\,\hat{F}_i \,+\,v\hat{\mathbbm{1}}
\end{align}
is an invariant such that all coefficients $\text{Tr} \dunderline{J}^i$ must be invariants for $i=1,\dots 8$.
Using (\ref{eq:decomposition_J_124567}-\ref{eq:decomposition_J_8}) we see that the trace for 
$i=1,2,4,5,6,7$ is trivially zero but for $i=3,8$ we get that $\text{Tr} \dunderline{J}^3=2 c_3$ and
$\text{Tr} \dunderline{J}^8=(2/\sqrt{3}) c_8$ must  be invariants. 

We note that only the $6$ coupling constants $(J_1,J_4,J_6)=(\xi_{12},\xi_{13},\xi_{23})$ and
$(K_1,K_4,K_6)=(\bar{\eta}_{12},\bar{\eta}_{13},\bar{\eta}_{23})$ are independent. This is consistent
with our general analysis in Section~\ref{sec:pms_N} where we showed that only the parameters
$\xi_{ll'}$ and $\bar{\eta}_{ll'}$ are needed. 

\begin{figure}
\centering
 \includegraphics[width=0.5\textwidth]{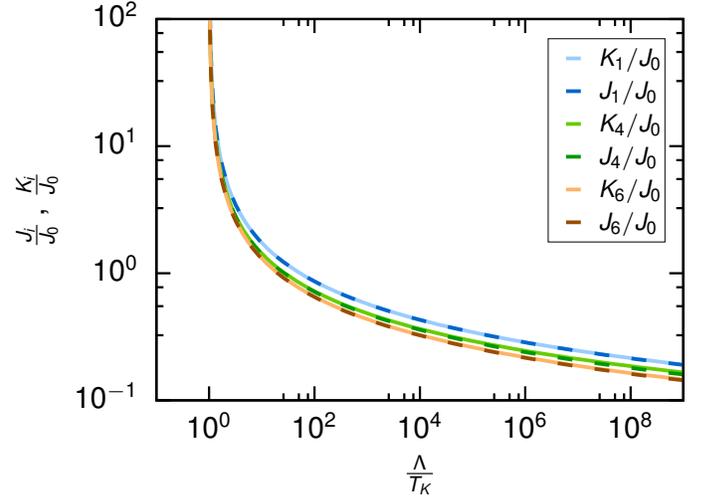}
 \caption{Flow of the poor man's scaling RG for the couplings with similiar initial values
$(J_1,J_4,J_6)(0) = (0.018235,0.015321,0.013784)J_0$,  
$(K_1,K_4,K_6)(0) = (0.018337,0.015924,0.013994)J_0$, $J_0=0.096510$ and $D=1000.0$.  
The couplings become degenerate at the Kondo scale $T_\text{K}$ and diverge.}
\label{fig:rg_flow}
\end{figure}

Since all coupling constants grow under the RG flow and diverge at $T_K$, the small invariants $c_3$,
$c_8$ and $v_l$ can be omitted from the effective interaction $V_\text{eff}$ defined
in (\ref{eq:V_eff_one_reservoir}). Inserting 
$\dunderline{\hat{J}}\,\approx\,\sum_{i=1}^8 \dunderline{J}^i\hat{F}_i$ from (\ref{eq:generators_dot_space})
and the decompositions (\ref{eq:decomposition_J_124567}-\ref{eq:decomposition_J_8}) we can write $V_\text{eff}$
in the compact form
\begin{align}
\label{eq:V_eff_f_F}
V_\text{eff}\,=\,2 \sum_{i=1}^8 \,J_i\,\hat{f}_i\,\hat{F}_i 
\,+\,2J_{38}\,(\hat{f}_8\hat{F}_3 + \hat{f}_3\hat{F}_8) \quad,
\end{align}
where we defined the reservoir $f$-spin operator as 
\begin{align}
\label{eq:f_spin}
\hat{f}_i\,=\,{1\over 2}\sum_{kk'}\uline{\tilde{a}}_k^\dagger\,\dunderline{\lambda}_i\,
\uline{\tilde{a}}_{k'}\quad.
\end{align}
The form (\ref{eq:V_eff_f_F}) exhibits very clearly how the reservoir $f$-spin couples
to the dot $F$-spin. There are three possible isospin pairs
formed by the up/down quark ($i=1,2$), the up/strange quark ($i=4,5$), 
or the down/strange quark ($i=6,7$), corresponding to the flavor pairs $l=1,2$, $l=1,3$ and
$l=2,3$, respectively. For each isospin pair we can define a transverse and
longitudinal coupling, denoted by $(J_1,K_1)$, $(J_4,K_4)$, and $(J_6,K_6)$, respectively, 
analog to the transverse and longitudinal Kondo couplings $(J_\perp,J_z)$ for a single spin $1/2$.
The three transverse couplings belong to the $6$ independent generators $\lambda_i$ for $i=1,2,4,5,6,7$.
Therefore, the effective interaction does not contain any transverse couplings between different
isospins of the reservoir and the dot but only the product $\hat{f}_i\,\hat{F}_i$ for $i=1,2,4,5,6,7$.
In contrast, the three longitudinal parts of the isospins are not independent. By convention
one chooses the longitudinal part of the up/down isospin (represented by $\lambda_3$) and
the sum over the longitudinal parts of the up/strange and down/strange isospins (represented by
the hypercharge generator $\sqrt{3}\lambda_8$) as basis for the two independent traceless matrices. 
Therefore, there is not only a longitudinal isospin coupling $J_3$ and a
hypercharge coupling $J_8$ but also a mixed coupling $J_{38}$ describing an
interaction of the longitudinal reservoir isospin with the hypercharge polarization of the dot and
vice versa. This picture naturally generalizes to arbitrary $N$ providing a physical interpretation
of the coupling constants $\xi_{ij}$ and $\bar{\eta}_{ij}$ in terms of the transverse and 
longitudinal couplings for the isospin formed by the two flavors $l=i,j$. 

Using (\ref{eq:rg_gamma}) and (\ref{eq:rg_eta_symmetric}) for $N=3$, we obtain the RG equations
\begin{align}
\label{eq:rg_J_1}
{dJ_1\over ds}\,&=\,2 J_1\,K_1\,+\,J_4\,J_6 \quad,\\ 
\label{eq:rg_J_4}
{dJ_4\over ds}\,&=\,2 J_4\,K_4\,+\,J_1\,J_6 \quad,\\ 
\label{eq:rg_J_6}
{dJ_6\over ds}\,&=\,2 J_6\,K_6\,+\,J_1\,J_4 \quad,\\ 
\label{eq:rg_K_1}
{dK_1\over ds}\,&=\,2 J_1^2\,+\,{1\over 2}(J_4^2\,+\,J_6^2) \quad,\\ 
\label{eq:rg_K_4}
{dK_4\over ds}\,&=\,2 J_4^2\,+\,{1\over 2}(J_1^2\,+\,J_6^2) \quad,\\ 
\label{eq:rg_K_6}
{dK_6\over ds}\,&=\,2 J_6^2\,+\,{1\over 2}(J_1^2\,+\,J_4^2) \quad,
\end{align} 
with the initial conditions at $s=0$ given by (\ref{eq:xi_eta_ll'})
\begin{align}
\label{eq:J_14_initial}
J_1(0)&={2\gamma_1\gamma_2\over D}\quad,\quad
J_4(0)={2\gamma_1\gamma_3\over D}\quad,\\
\label{eq:J_6_K_1_initial}
J_6(0)&={2\gamma_2\gamma_3\over D}\quad,\quad
K_1(0)={\gamma^2_1+\gamma^2_2\over D}\quad,\\
\label{eq:K_46_initial}
K_4(0)&={\gamma_1^2+\gamma_3^2\over D}\quad,\quad
K_6(0)={\gamma_2^2+\gamma^2_3\over D}\quad.
\end{align}
A numerical study of these RG equations shows that independent of the initial conditions all couplings become equal
during the RG flow and diverge at some low-energy scale $T_K$, in agreement with (\ref{eq:xi_eta_fixed_point}). 
Using (\ref{eq:J_K_12}-\ref{eq:J_38}) this means that all $J_i=J/2$ become the same 
for $i=1,\dots,8$ and the mixed coupling $J_{38}$ scales to zero. Thus, at the fixed point the 
effective interaction can be written in the isotropic and $SU(3)$-invariant form
\begin{align}
\label{eq:V_eff_SU3}
V_\text{eff}\,=\,J\,\sum_{i=1}^8 \,\hat{f}_i\,\hat{F}_i \quad,
\end{align}
which is identical with (\ref{eq:V_fixed_point}). Applying the analog scheme to an arbitrary number $N$ 
of dot levels we obtain at the fixed point the same result, one just has to sum in (\ref{eq:V_eff_SU3}) 
over all generators of $SU(N)$. Fig.~\ref{fig:rg_flow} shows an example for the RG flow where the
longitudinal and tranverse couplings $K_i\approx J_i$ are initially nearly the same but different for each 
$i=1,4,6$.

\begin{figure}
\centering
 \includegraphics[width=0.5\textwidth]{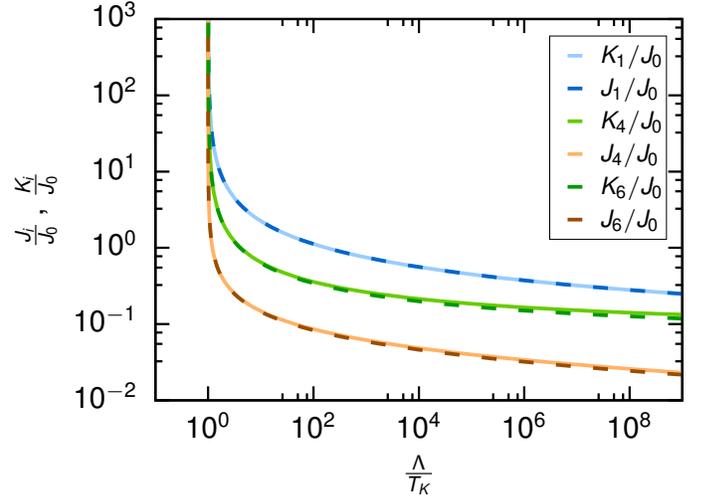}
 \caption{Flow of the poor man's scaling RG for the couplings for $J_1(0) \gg J_4(0), J_6(0)$ with 
$(J_1,J_4,J_6)(0) = (0.0239873,0.0022176,0.0020878)J_0$, 
$(K_1,K_4,K_6)(0) = (0.0240310,0.0128358,0.0113882)J_0$, $J_0=0.0965099$, $D=1000.0$.
Each of the coupling pairs $(J_1,K_1)$, $(J_4,J_6)$ and $(K_4,K_6)$ are quasi degenerate 
for the main part of the RG flow before all couplings obtain the same value at $T_\text{K}$.}
\label{fig:isospin_picture}
\end{figure}
To obtain a feeling for the nature of the strong-coupling ground state, we assume a two-site model 
with Hamiltonian \eqref{eq:V_eff_SU3}. In particular, we consider a tight-binding model for the reservoir 
and the two sites are the dot and the first site of the reservoir (i.e. the one that couples to the dot), 
respectively, while the other reservoir sites are not taken into account. The crucial point about 
determining the ground state lies in choosing the approrpriate representation for the eigenstates of 
the $SU(3)$-symmetric interaction in \eqref{eq:V_eff_SU3}. The $SU(3)$ group has two fundamental 
representations \cite{su3_group}, which we denote by the multiplett notation $[3]$ and $[\overline{3}]$. 
We represent the eigenstates of the dot in the representation $[3]$ where the $F$-spin components are 
$\hat{F}_i={1 \over 2} \hat{\lambda}_i$. Denoting the states by the quark flavors $l=1,2,3=u,d,s$, we have
\begin{align}
\label{eq:up_quark_state}
\ket{u} \,&=\, \ket{{1\over 2}, {1 \over 3}} \quad,\\
\label{eq:down_quark_state}
\ket{d} \,&=\, \ket{-{1\over 2}, {1 \over 3}} \quad,\\
\label{eq:strange_quark_state}
\ket{s} \,&=\, \ket{0, -{2 \over 3}} \quad,
\end{align}
where the states on the r.h.s. are the eigenstates of $\hat{F}_3$ and $\hat{F}_8$ and the 
first (second) quantum number in the label is the corresponding eigenvalue of 
$\hat{F}_3$ (${2 \over \sqrt{3}} \hat{F}_8$). Therefore, we refer to these eigenvalues as 
isospin (hypercharge) quantum numbers. Choosing the same representation for the first site in the 
reservoir is not useful since the states of the composite system are part of either the sextet $[6]$ 
or the triplet $[\overline{3}]$ due to $[3] \otimes [3]= [6] \oplus [\overline{3}]$ \cite{su3_group}. 
Such a representation is not suitable since the system has a distinct non-degenerate ground state. 
Instead, we represent the first site of the reservoir with $[\overline{3}]$ and obtain  
$[3] \otimes [\overline{3}]= [8] \oplus [1]$ where all but one state of the two-site system 
form an octet together with the remaining state being a unique singlet state. $[\overline{3}]$ is the 
complex conjugate representation of $[3]$ and has  therefore the generators 
$\hat{f}_i= - {1 \over 2} \hat{\lambda}^*_i$. Consequently, we label the states of the second site 
with the anti-quark flavor $\overline{l}=1,2,3=\overline{u},\overline{d},\overline{s}$ and get
\begin{align}
\label{eq:up_antiquark_state}
\ket{\overline{u}} \,&=\, \ket{-{1\over 2},-{1 \over 3}} \quad,\\
\label{eq:down_antiquark_state}
\ket{\overline{d}} \,&=\, \ket{{1\over 2}, -{1 \over 3}} \quad,\\
\label{eq:strange_antiquark_state}
\ket{\overline{s}} \,&=\, \ket{0, {2 \over 3}} \quad.
\end{align}
In this basis, the operators $\hat{\lambda}^*_i$ have the same matrix representation as the 
Gell-Mann matrices $\hat{\lambda}_i$.

Indeed, we will show in appendix \ref{app:ground_state_of_fixed_point_model} that the singlet state
\begin{align}
\label{eq:ground_state_fixed_point_model}
\ket{\text{gs}} \,=\, {1 \over \sqrt{3}} \left( \ket{u \overline{u}} 
+ \ket{d \overline{d}} + \ket{s \overline{s}} \right)
\end{align}
is the ground state with energy $E_\text{gs}=-{4 \over 3} J$ while the octet states are degenerate
with energy $E_8={1 \over 6}J$. Since $\ket{l \overline{l}}=\ket{l} \otimes \ket{\overline{l}}$ 
it is straightforward to define the reduced dot density matrix
\begin{align}
\label{eq:reduced_dot_density_matrix}
\hat{\rho} \,=\, \sum_{\overline{l}=\overline{u},\overline{d},\overline{s}} \bra{\overline{l}} 
\left( \ket{\text{gs}} \bra{\text{gs}} \right) \ket{\overline{l}} 
\,=\, {1 \over 3} \mathbbm{\hat{1}}\quad,
\end{align}
which yields $n_l={1 \over 3}$ in perfect agreement with the NRG analysis in section \ref{sec:nrg}.

Together with the $SU(3)$-symmetric interaction term, the outcome \eqref{eq:ground_state_fixed_point_model} 
motivates the term "quantum fluctuations" for the significant physical processes in the fixed point model. 
The ground state is a symmetric linear combination of bound states with quark-antiquark-flavor. This is 
in accordance with the observation that no free quarks exist, i.e. they always gather to form a particle 
with integer electric charge. The interaction term \eqref{eq:V_eff_SU3} preserves this since the 
fluctuation terms ($i=1,2,4,5,6,7$) always annihilate a quark-antiquark-pair while creating a 
different quark-antiquark-bound state simultaneously. Furthermore, we will discuss in 
Appendix~\ref{app:ground_state_of_fixed_point_model} that the eigenstates of \eqref{eq:V_eff_SU3} 
are identical to those of the quark model for light pseudoscalar mesons\cite{quarks}.

In this context, choosing $J_1 \approx K_1 \gg J_4 \approx J_6$ and $K_4 \approx K_6$
for the initial values reveals a nice physical picture in terms of the isospin of the up and down
quark. Fig.~\ref{fig:isospin_picture} shows that in the whole regime from weak to intermediate coupling 
the couplings stay approximately degenerate with $J_1 \approx K_1$, $J_4 \approx J_6$ and $K_4 \approx K_6$. 
Here, the model exhibits an approximated $SU(2)$-symmetry for the isospin with an isotropic isospin 
coupling $J_I = {1 \over 2} (J_1 + K_1) \gg \vert J_1 - K_1 \vert$. Furthermore, the interaction of 
isospin and hypercharge degrees of freedom disentangle in leading order since $J_{38} \ll J_3, J_8$. 
In the same way, $J_4 \approx J_6$ characterizes transitions between states differing in the hypercharge 
quantum number, compare with (\ref{eq:up_quark_state}-\ref{eq:strange_quark_state}). In total, we find an 
isotropic isospin model where the presence of the third level (strange quark) mainly results in a 
potential scattering ($J_8 \sim J_I$) for the isospin with suppressed transitions to states with different 
hypercharge ($J_4, J_6 \ll J_I$). However, finally the RG flow approaches the generic $SU(3)$-symmetric fixed point 
on the Kondo scale $T_\text{K}$ also in this case.

\section{NRG analysis in equilibrium}
\label{sec:nrg}

\begin{figure*}
  \centering
  ~~
  \includegraphics[width=0.425\textwidth]{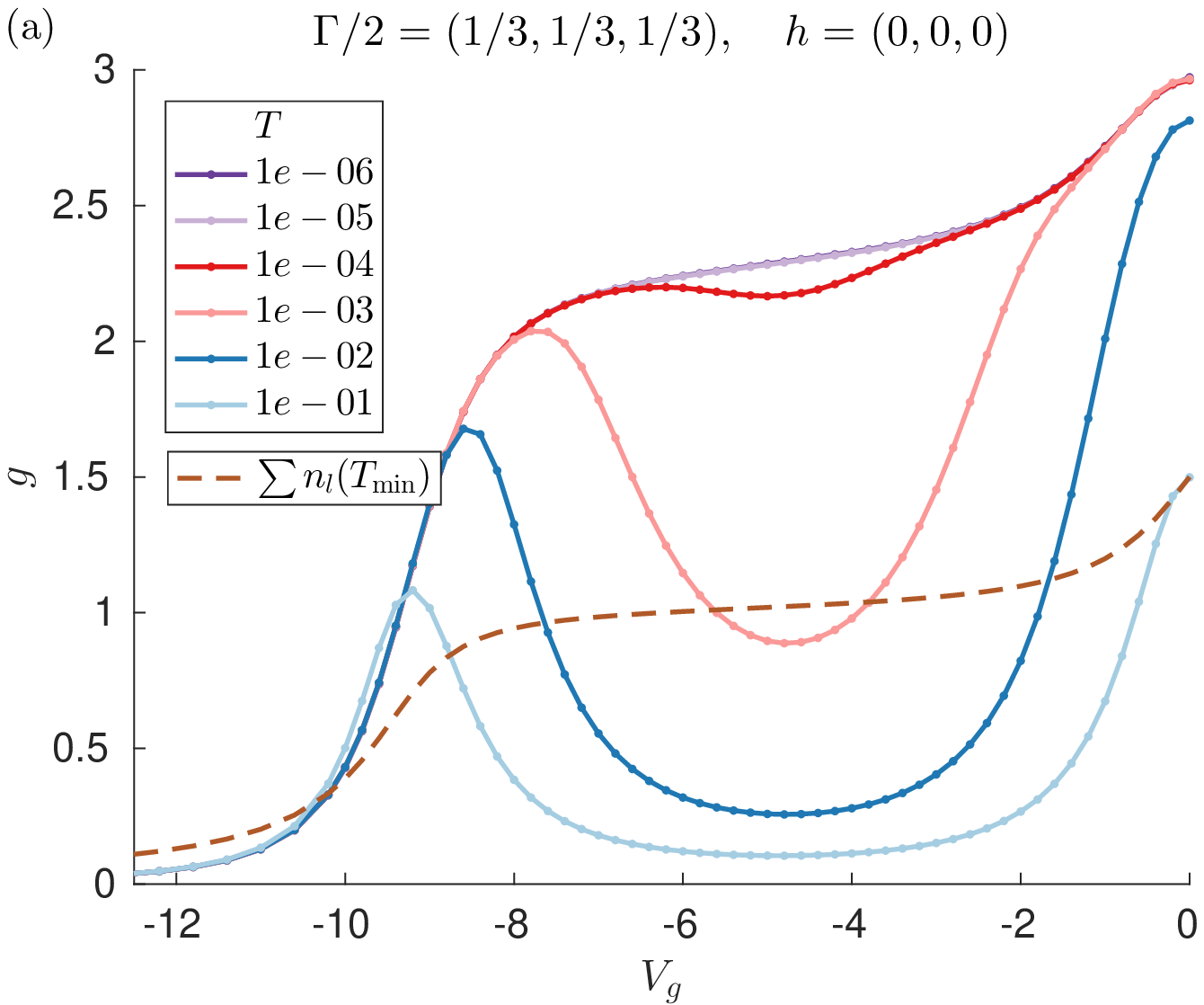}
  \hfill
  \includegraphics[width=0.425\textwidth]{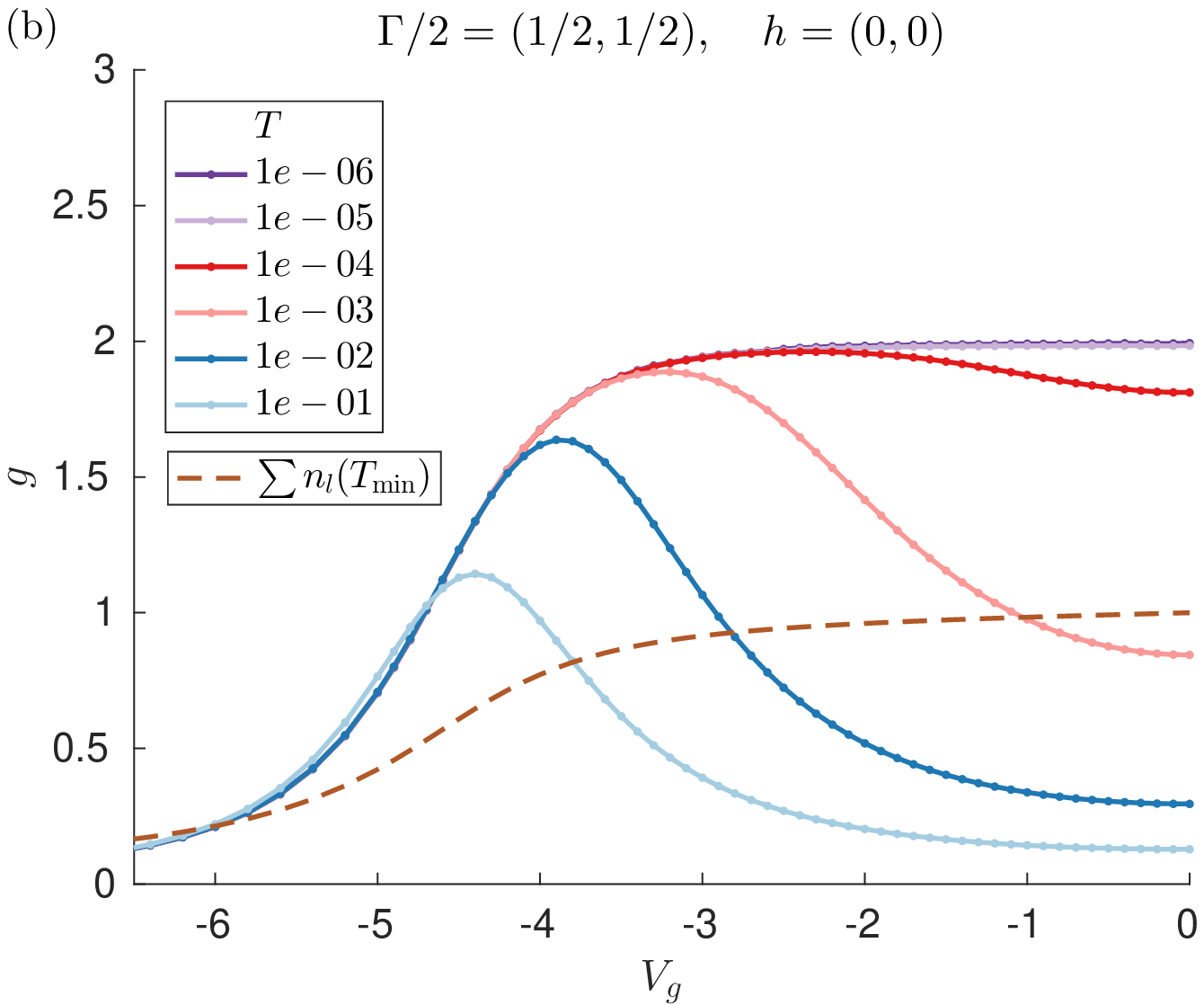}
  ~~
  \caption{Gate voltage dependence of the conductance at various temperatures for (a) $N=3$ 
    and (b) $N=2$ at the $SU(N)$-symmetric point where all $\Gamma_l$ are the same and all $h_l=0$. 
    By p-h symmetry the curves can be mirrored along $V_g=0$. Also shown is the occupation $n_l$ at
    the lowest temperature where the Friedel sum rule \eqref{eq:friedel_sum_rule} is fulfilled.}
  \label{fig:SU(N)_sym}
\end{figure*}
In Section~\ref{sec:pms_quarks} we have shown for a $3$-level quantum dot in the cotunneling regime 
that the generic fixed point model is an $SU(3)$-invariant isotropic effective interaction 
(\ref{eq:V_eff_SU3}) between the $F$-spins of the reservoir and the dot. This holds  
for the equilibrium case where all reservoirs can be taken together to a single reservoir and it 
requires also $SU(3)$-symmetry of the dot. This means that the dot parameters $h_{ll'}$ have to be 
adjusted appropriately (including renormalizations arising from the coupling to the reservoir) 
such that the populations of all dot states are the same $n_l=\langle c^\dagger_l c_l\rangle=1/3$. 
The aim of this section is to confirm that in equilibrium the $SU(3)$-symmetric fixed point can be
established independent of the tunneling matrix by an adjustment of the dot parameters. To this end we
use the numerically exact NRG method \cite{Bulla2008} and analyse the linear conductance $G$ for $N=3$
and two reservoirs ($\alpha=L,R$) for the case of proportional couplings 
$\dunderline{\Gamma}_\alpha=x_\alpha\dunderline{\Gamma}$ where $G$ can be calculated from
(\ref{eq:conductance_tensor}) and (\ref{eq:transmission})
\begin{align}
\label{eq:G_two_reservoirs}
g\,=\,G/G_0\,=\,-{\pi\over 2} \int d\omega \,
\text{Tr} \,\dunderline{\Gamma}\,\dunderline{\rho}(\omega)\,f'(\omega) \quad,
\end{align}
with the dimensionless conductance $g$ in units of $G_0=(e^2/h)/(4x_L x_R)$.
As explained in Section~\ref{sec:tunneling_model} the equilibrium spectral density 
$\dunderline{\rho}(\omega)$ depends only on the total hybridization matrix $\dunderline{\Gamma}$,
i.e. we can use a unitary transformation of the dot states such that this matrix is diagonal (see
(\ref{eq:gamma_total_diagonal})) and the spectral density in this basis depends only on 
the eigenvalues $\Gamma_l=2\pi\gamma_l^2$. In this case the linear conductance (\ref{eq:G_two_reservoirs})
can be written as
\begin{align}
\label{eq:G_two_reservoirs_diagonal}
g\,=\,-{\pi\over 2} \int d\omega \,\sum_l \,\Gamma_l \,{\rho}_{ll}(\omega) \,f'(\omega) \quad.
\end{align}
In the new dot basis we assume for simplicity that the dot Hamiltonian contains only diagonal elements
\begin{align}
\label{eq:dot_H_diagonal}
H\,=\,\sum_l h_l c^\dagger_l c_l \quad.
\end{align}
Other cases with nondiagonal elements $h_{ll'}$ can also be studied but are of no interest
because they just destroy $SU(3)$-symmetry of the dot and drive the system away from the fixed point model.
Here, we are interested in a systematic study how, for {\it arbitrary} tunneling parameters $\Gamma_l$, 
$SU(3)$-symmetry can be restored by tuning the level positions $h_l$ appropriately. In addition we will
also study the dependence of the $SU(3)$-Kondo temperature $T_K^{(3)}$ as function of $\Gamma_l$ and compare
it to the corresponding $SU(2)$-Kondo temperature $T_K^{(2)}$, where only two levels contribute to transport.  
This analysis goes beyond the one of Ref.~\onlinecite{lopez_etal_13} which has concentrated on the 
linear conductance for the $SU(3)$-symmetric case (i.e. all $\Gamma_l$ are the same and $h_l=0$) and
the destruction of $SU(3)$-symmetry by different $\Gamma_l$ or finite values for $h_{ll'}$. As a 
signature of $SU(3)$-symmetry we take the Friedel sum rule (used also in Ref.~\onlinecite{Moca2012,lopez_etal_13})
\begin{align}
\label{eq:friedel_sum_rule}
g\,=\,\sum_l \sin^2(\pi n_l) \quad,
\end{align}
which holds exactly at zero temperature and gives the value $g=2.25$ for equal populations
$n_l=1/3$ corresponding to the $SU(3)$-symmetric fixed point. The occupations $n_l$ can be calculated from
the spectral density via $n_l=\int d\omega \rho_{ll}(\omega)f(\omega)$. For the parameters in all figures we use 
\begin{align}
\label{eq:parameters}
{1\over 2}\sum_l\Gamma_l\,=\,1 \quad,\quad U\,=\,10 \quad,\quad W\,=\,10^4 \quad,
\end{align}
where $2W$ denotes the width of a flat d.o.s. of the reservoirs (i.e. $|\omega|<W$ for the integral in 
(\ref{eq:G_two_reservoirs_diagonal})). 

\begin{figure}
  \centering
  \includegraphics[width=0.425\textwidth]{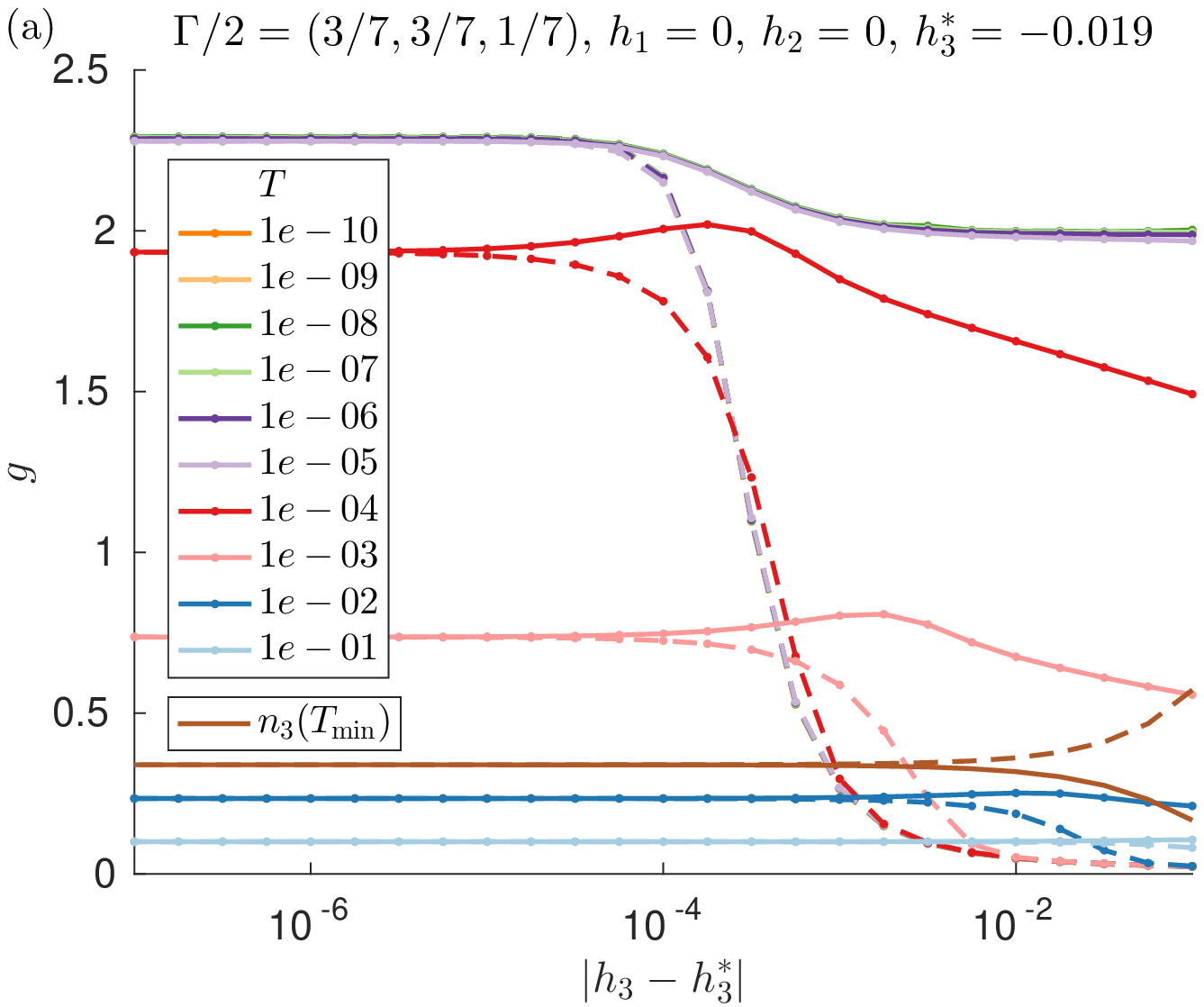}\\
  \vspace{1em}
  \includegraphics[width=0.425\textwidth]{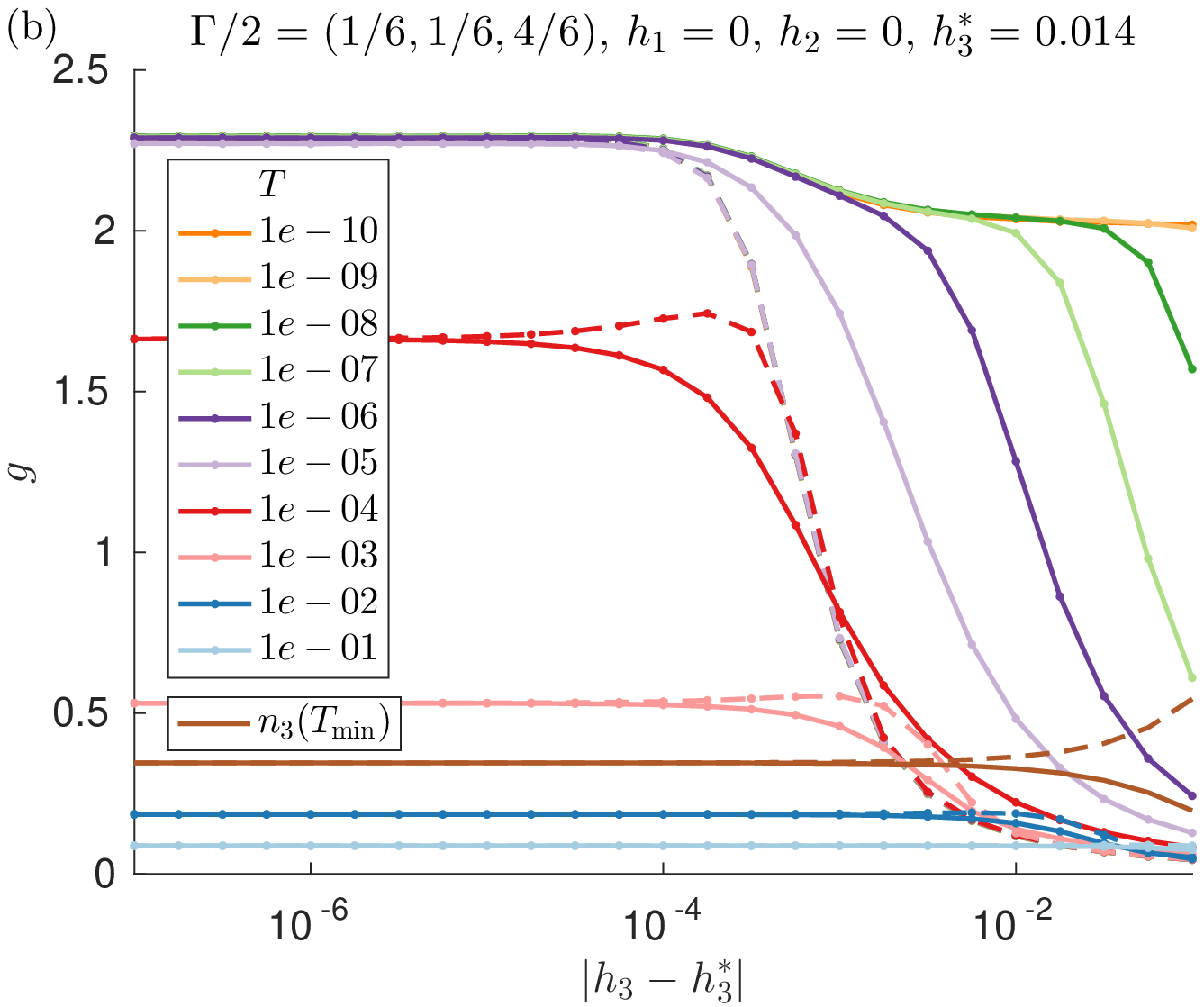}
  \caption{Conductance at fixed $V_g=-U/2$ and $h_1=h_2=0$ for (a) $\Gamma_1=\Gamma_2 > \Gamma_3$ and
    (b) $\Gamma_1=\Gamma_2 < \Gamma_3$ as function of $|h_3-h_3^*|$ for various temperatures. 
    We distinguish the case $h_3 > h_3^*$ (solid lines) from the case $h_3 < h_3^*$ (dashed lines), 
    where $h_3^*$ is the optimized value at which $SU(3)$-symmetry is restored.}
  \label{fig:gam1=gam2_g(h_3)}
\end{figure}
The calculations are performed using the full-density-matrix NRG \cite{Weichselbaum2007-2012}, 
where we exploit either the individual charge conservation or the full $SU(N)$ symmetry by 
means of the QSpace tensor library developed by A. Weichselbaum \cite{Weichselbaum2012a}. 
For the final results we employ a discretization parameter of $\Lambda=3$, and we keep states up to a 
rescaled energy of $E_{\textrm{trunc}}=9$ and maximal number $N_{\textrm{keep}}$ during the NRG iteration. 
In the calculations without $SU(N)$ symmetry we set $N_{\textrm{keep}}=8000$. In the $SU(N)$-symmetric cases 
we can further increase the precision to very high level and explicitly confirm that results for $g$ are 
converged up to $1\%$ and results for $n_l$ are converged up to $10^{-6}$ with respect to the numerical 
parameters. Note that in many calculations we optimize the level positions to achieve equal occupation of 
certain levels. Since the values of such optimized level positions $h_l^*$ depend on the discretization of 
the bath, we refrain from using $z$-averaging \cite{Bulla2008}. Finally, we need not broaden the NRG data as 
the computation of both $g$ and $n_l$ requires only discrete spectral weights.

To set the scene we show in Fig.~\ref{fig:SU(N)_sym} known curves for the conductance depending on gate 
voltage and temperature 
in the $SU(N)$-symmetric cases for $N=2,3$, where all $\Gamma_l$ are the same and all $h_l=0$. We find
converged, plateau-like features when decreasing $T$ below the Kondo temperature $T_K$ in the cotunneling 
regime of a singly occupied dot. Note that $n_l$ shows a very weak dependence on temperature in this regime and, 
at $T<T_K$, the Friedel sum rule \eqref{eq:friedel_sum_rule} is fulfilled. Furthermore, we find
that the Kondo temperatures $T_K^{(N)}$ are similiar for $N=2$ and $N=3$
(recall that $\sum_l \Gamma_l$ is fixed). In contrast, the p-h-symmetric point 
$V_g=0$ corresponds to very different physics for the two cases, since for $N=3$ there are strong 
charge fluctuations due to $E_1=E_2$, whereas for $N=2$ spin fluctuations
dominate. Therefore, at $V_g=0$, the relevant low-energy scale is the hybridization $\Gamma_l$ for $N=3$
\cite{lopez_etal_13} and the Kondo temperature for $N=2$. 

\begin{figure}
  \centering
  \includegraphics[width=0.425\textwidth]{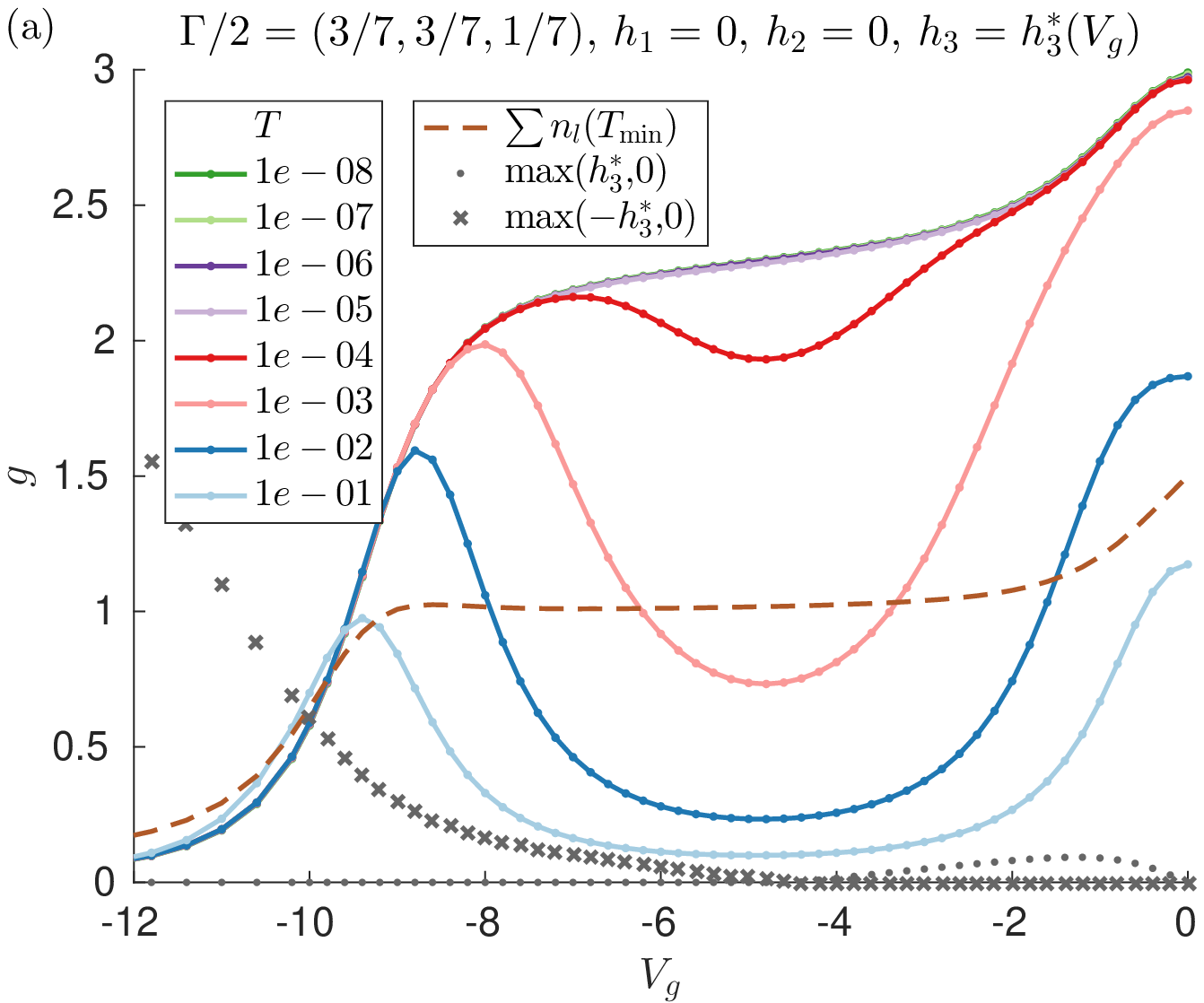}\\
  \vspace{1em}
  \includegraphics[width=0.425\textwidth]{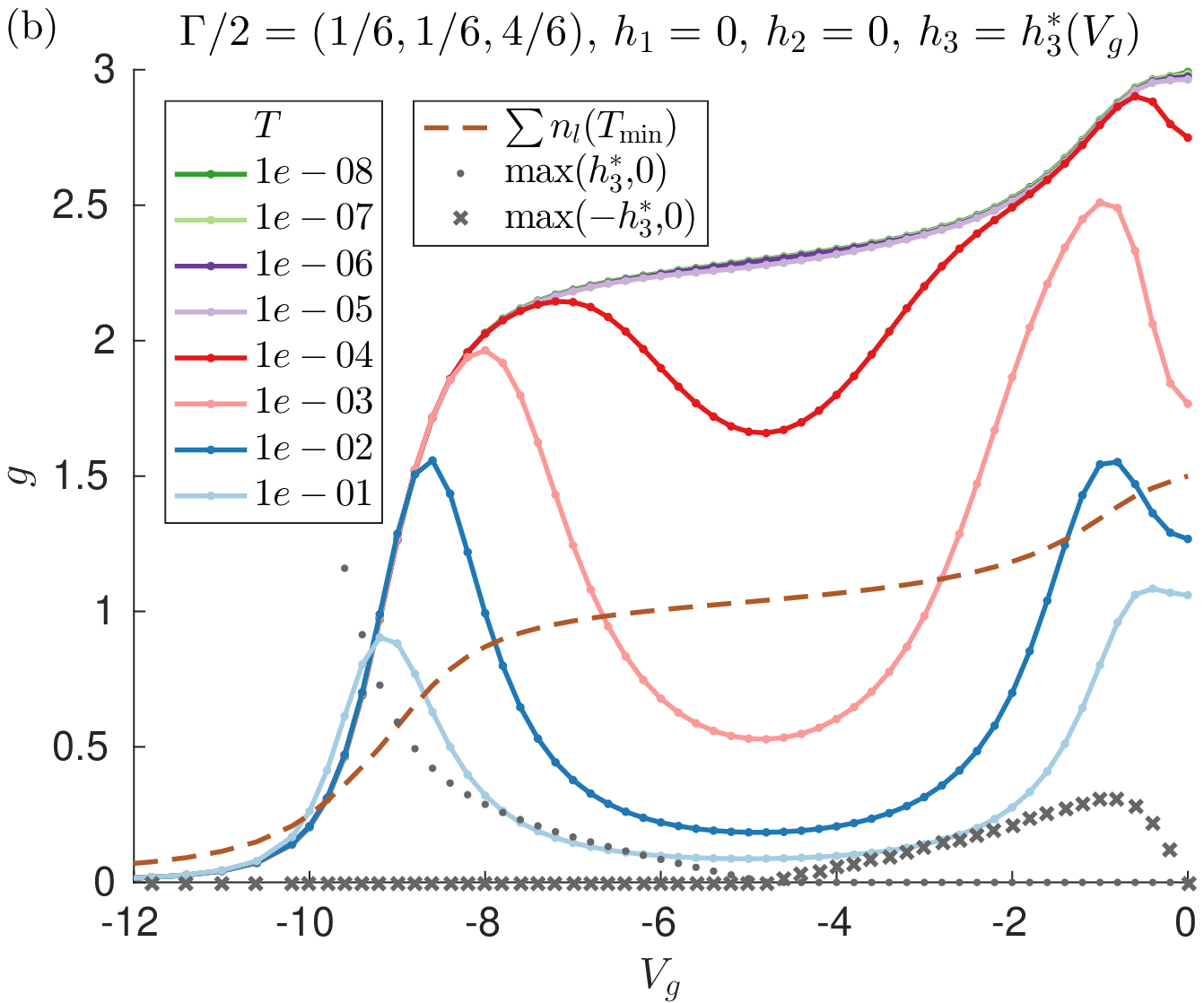}
  \caption{Conductance for $h_1=h_2=0$ and (a) $\Gamma_1=\Gamma_2 > \Gamma_3$ and
    (b) $\Gamma_1=\Gamma_2 < \Gamma_3$ as function of gate voltage for various temperatures. 
    For each value of the gate voltage $h_3=h_3^*(V_g)$ is optimized such that the populations
    of the three states are the same at zero temperature.}
  \label{fig:gam1=gam2_g(Vg)}
\end{figure}
Next we study the case $\Gamma_1=\Gamma_2\ne\Gamma_3$ and $h_1=h_2=0$. 
In this case the different tunneling couplings lead to a different renormalization of $h_3$ of 
$O(\Gamma_1\Gamma_3/U)$ relativ to $h_{1/2}$. Therefore, $h_1=h_2=h_3=0$ is {\it not} 
the $SU(3)$-symmetric point and the level position $h_3$ has to be adjusted appropriately to recover 
equal populations of the states and conductance $g=2.25$ at zero temperature. Calling this optimized
value $h_3^*$ we show in Fig.~\ref{fig:gam1=gam2_g(h_3)} the conductance as function of $|h_3-h_3^*|$.
For temperatures $T<T_K^{(3)}$ we see that the conductance reaches the $SU(3)$-symmetric value $g=2.25$ for
$|h_3-h_3^*|\sim T_K^{(3)}$ as expected. The Kondo temperature $T_K^{(3)}$ does not depend strongly on the
value of $\Gamma_3$ and is nearly the same for $\Gamma_3<\Gamma_{1/2}$ 
(Fig.~\ref{fig:gam1=gam2_g(h_3)}(a)) and $\Gamma_3>\Gamma_{1/2}$ (Fig.~\ref{fig:gam1=gam2_g(h_3)}(b)). 
For $|h_3-h_3^*|>T_K^{(3)}$ and $h_3>h_3^*$ (solid lines in Fig.~\ref{fig:gam1=gam2_g(h_3)}) we see 
that the $SU(2)$-Kondo effect with $g=2$ appears at low enough 
temperatures $T<T_K^{(2)}$. Whereas $T_K^{(2)}\approx T_K^{(3)}$ for relatively small $\Gamma_3<\Gamma_{1/2}$, we find 
that $T_K^{(2)}<T_K^{(3)}$ for $\Gamma_3>\Gamma_{1/2}$. The latter can be explained by the fact that 
the two levels $l=1,2$ form the $SU(2)$-Kondo effect and therefore $T_K^{(2)}$ decreases if the 
coupling to these two levels $\Gamma_1, \Gamma_2$ is lowered. In contrast, when all three levels contribute to the
$SU(3)$-Kondo effect, we have a total coupling of $\sum_l \Gamma_l/2=1$ and find that the relative 
distribution of the $\Gamma_l$ influences $T_K^{(3)}$ only weakly. Furthermore, in the regime where 
the $SU(2)$-Kondo effect occurs we see a strong difference when moving over from $h_3>h_3^*$ to $h_3<h_3^*$ 
(dashed lines in Fig.~\ref{fig:gam1=gam2_g(h_3)}) since then level $3$ forms the ground state and thus the 
Kondo effect is much weaker compared to the case when the two levels $l=1,2$ are lower in
energy. In the regime of the $SU(3)$-Kondo effect it is hardly relevant whether level $3$ approaches the
other two levels from above or below.  
\begin{figure}
  \centering
  \includegraphics[width=0.475\textwidth,trim={1.4cm 0.3cm 4.85cm 1.1cm},clip]{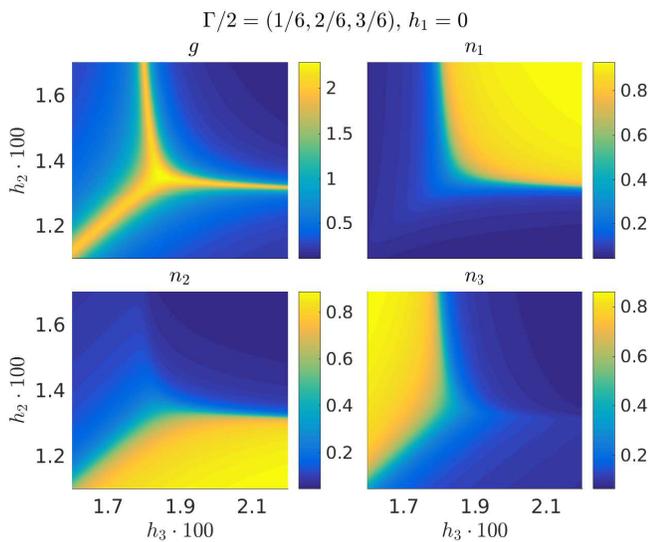}
  \caption{Conductance and level occupations as functions of $h_2$ and $h_3$ for 
    $\Gamma_1<\Gamma_2<\Gamma_3$, $h_1=0$, $V_g=-U/2$, and $T=10^{-10}$.}  
  \label{fig:g_h2_h3}
\end{figure}

In Fig.~\ref{fig:gam1=gam2_g(Vg)} we show the conductance as function of the gate voltage again for
$h_1=h_2=0$ and the two cases $\Gamma_1=\Gamma_2\gtrless\Gamma_3$ as in Fig.~\ref{fig:gam1=gam2_g(h_3)} but at
each value of the gate voltage we choose the optimized value $h_3=h_3^*(V_g)$ for which the populations
of the three states are the same at zero temperature. As in Fig.~\ref{fig:gam1=gam2_g(h_3)} we confirm 
that $T_K^{(3)}$ depends only weakly on $\Gamma_3$ but the overall tendency is that $T_k^{(3)}$ decreases when 
increasing $|\Gamma_{1/2}-\Gamma_3|$. At the p-h symmetric point $V_g=0$, the situation is completely
different since charge fluctuations dominate for $N=3$. Therefore, the conductance around $V_g=0$ 
depends strongly on the relative distribution of the $\Gamma_l$. In fact, comparing various cases we find 
that the conductance at $V_g=0$ (where also $h_3^*=0$) decreases monotonously when increasing the variance 
of the couplings $\Gamma_l$. At large variance as in Fig.~\ref{fig:gam1=gam2_g(Vg)}(b), $g$ around $V_g=0$ 
is strongly surpressed. In contrast, in the cotunneling regime $V_g\approx-U/2$ the conductance is 
rather insensitive to the distribution of the $\Gamma_l$. The combination of these phenomena leads 
to a surprising shape of the curve $g(V_g)$ which exhibits a local minimum at the p-h symmetric point 
for intermediate temperatures.

Finally, we consider in Fig.~\ref{fig:g_h2_h3} three different hybridizations $\Gamma_1<\Gamma_2<\Gamma_3$ 
and tune $h_2$ and $h_3$ at fixed $h_1=0$, $V_g=-U/2$, and $T=10^{-10}$. From the plots of the occupations 
$n_l$ we can easily distinguish three regions where only one level is involved. At the intersections 
of two such regions we observe a two-level Kondo effect with conductance $g=2$. The widths 
of these intersections in the $h_2$-$h_3$ plane define three different Kondo temperatures $T_K^{(2)}$
which are ordered according to the size of the corresponding hybridizations 
$\Gamma_1+\Gamma_2<\Gamma_1+\Gamma_3<\Gamma_2+\Gamma_3$. In the center, where all ``one-level sections'' 
intersect, we observe a wide region of a three-level Kondo effect with conductance $g=2.25$.
The corresponding Kondo temperature $T_K^{(3)}$ is of the same order as the maximum of the three
two-level Kondo temperatures. 

In summary, we find that for any kind of (diagonal) hybridization,
whether with no, two, or three identical elements, we can find carefully optimized
level positions (and low enough temperatures) to observe the behavior known from the 
$SU(3)$-symmetric quantum dot. For other hybridizations with two identical hybridization elements or, 
again, optimized level positions we can also reproduce the behavior of a two-level 
Kondo effect such that one level is (effectively) excluded. For arbitrary $\Gamma_l$ and 
$h_l$ (corresponding to most parts of a version of Fig.~\ref{fig:g_h2_h3} zoomed out) 
the typical behavior is that of the single, (effectively) lowest-lying level.

\section{Nonequilibrium fixed point model}
\label{sec:nonequilibrium_fixed_point}

The aim of this section is to analyse the {\it nonequilibrium} properties of the system for $N=3$ in the
perturbative regime where the cutoff scale $\Lambda_c$ defined by (\ref{eq:Lambda_c}) is much larger
than the Kondo temperature $\Lambda_c\gg T_K$. Most importantly, as already emphasized several times 
in the previous sections, even if the fixed point model (\ref{eq:V_eff_SU3}) is reached at 
scale $\Lambda_c$ (which will be the case if we take the formal scaling limit defined by 
(\ref{eq:scaling_limit})), it is essentially {\it not} $SU(3)$-invariant if the chemical
potentials of all reservoirs are different. This leads to new interesting nonequilibrium fixed point
models similiar to the ones discussed in Ref.~\onlinecite{goettel_reininghaus_schoeller_15} for the $N=2$
case which show a completely different behaviour of physical observables like the magnetization or the current
compared to the $SU(N)$-symmetric Kondo model. Moreover, in practical situations the initial cutoff
$D\sim E_c$ is fixed leading to deviations from the fixed point model. Therefore, the aim of this section
is to analyse the perturbative effects of the full effective interaction on physical observables and to identify a
{\it smoking gun} for the fixed point model together with a parameter measuring the distance from this
fixed point. 

\subsection{Golden rule approach}
\label{sec:golden_rule}

We start from the effective interaction in the form (\ref{eq:V_eff}) in terms of the original reservoir 
field operators $a_{\alpha l k}$. Inserting (\ref{eq:generators_dot_space}-\ref{eq:decomposition_J_8}) 
and leaving out all small terms $\sim v_l, c_3, c_8$, we obtain
\begin{align}
\label{eq:V_eff_RG}
V_\text{eff}\,=\,\sum_{\alpha\alpha'}\sum_{kk'}\,:\uline{a}^\dagger_{\alpha k}\,
\,\dunderline{V}_\alpha\,\dunderline{\hat{J}}\,\dunderline{V}_{\alpha'}^\dagger\,\uline{a}_{\alpha'k'}\quad,
\end{align}
with
\begin{align}
\label{eq:J_RG}
\dunderline{\hat{J}}\,&\approx\,\sum_i\,\dunderline{J}^i\,\hat{F}_i \\
\label{eq:J_i_RG}
\dunderline{J}^i\,&=\,J_i\,\dunderline{\lambda}_i
\,+\,J_{38}(\delta_{i3}\,\dunderline{\lambda}_8\,+\,\delta_{i8}\,\dunderline{\lambda}_3)\quad.
\end{align}

The total Hamiltonian is given by $H_\text{tot}=H_\text{res} + H_\text{dot} + V_\text{eff}$, with a unity 
d.o.s. in the reservoirs and the dot Hamiltonian $H_\text{dot}=\sum_{ll'}h_{ll'}|l\rangle\langle l'|$ in 
the $1$-particle subspace. To apply golden rule we first diagonalize the dot Hamiltonian by
a unitary transformation $\hat{U}$ such that 
\begin{align}
\label{eq:H_diagonal}
\tilde{H}_\text{dot}\,=\,\hat{U}^\dagger H_\text{dot} \hat{U} \,=\, \sum_l \epsilon_l |l\rangle\langle l| \quad.
\end{align}
The golden rule rate for a transition from $l'\rightarrow l$ in the diagonalized basis is then given by 
\begin{align}
\nonumber
\Gamma_{l'\rightarrow l} \,&=\,
2\pi \sum_{rr'}|\langle l r|\hat{U}^\dagger V_\text{eff} \hat{U} |l'r'\rangle|^2 
\langle r'|\rho_{\text{res}}|r'\rangle\,\cdot \\
\label{eq:golden_rule_general}
&\hspace{2cm}
\cdot\,\delta(\epsilon_l + E_r - \epsilon_{l'} - E_{r'}) \quad,
\end{align}
where $|r\rangle$ denote the many-particle states of the reservoirs with energy $E_r$ and
$\rho_\text{res}=\prod_\alpha \rho_\text{res}^\alpha$ is the product of the grandcanonical 
distributions of the reservoirs. Inserting the effective interaction (\ref{eq:V_eff_RG}) we find
\begin{widetext}
\begin{align}
\label{eq:golden_rule_Fermi_function}
\Gamma_{l'\rightarrow l} \,=\, 2\pi \sum_{\alpha\alpha'}\int d\omega \int d\omega' (1-f_\alpha(\omega))f_{\alpha'}(\omega')
\delta(\epsilon_l-\epsilon_{l'} +\omega + \mu_\alpha - \omega' - \mu_{\alpha'})
\sum_{l_1 l_1^\prime}
|\langle l|\hat{U}^\dagger (\dunderline{V}_\alpha\,\dunderline{\hat{J}}\,
\dunderline{V}_{\alpha'}^\dagger)_{l_1 l_1^\prime} \hat{U}|l'\rangle|^2\quad.
\end{align}
\end{widetext}
At zero temperature we get
\begin{align}
\nonumber
\Gamma_{l'\rightarrow l} \,&=\, 2\pi \sum_{\alpha\alpha'} w(\epsilon_l-\epsilon_{l'}+\mu_\alpha - \mu_{\alpha'})\,\cdot\\
\label{eq:rate}
&\hspace{2cm}
\cdot\,\sum_{l_1 l_1^\prime}|\langle l|\hat{U}^\dagger (\dunderline{V}_\alpha\,\dunderline{\hat{J}}\,
\dunderline{V}_{\alpha'}^\dagger)_{l_1 l_1^\prime} \hat{U}|l'\rangle|^2\quad,
\end{align}
with $w(x)=|x|\theta(x)$. Here, $|\epsilon_l-\epsilon_{l'}+\mu_\alpha - \mu_{\alpha'}|$
is just the available energy phase space in the reservoirs for the energy gain 
$\epsilon_{l'}-\epsilon_l+\mu_{\alpha'} - \mu_\alpha > 0$. Inserting 
(\ref{eq:J_RG}) we can write the golden rate in the compact form
\begin{align}
\nonumber
\Gamma_{l'\rightarrow l}\,&=\,2\pi \sum_{\alpha\alpha'} w(\epsilon_l-\epsilon_{l'}+\mu_\alpha - \mu_{\alpha'})\,\cdot\\
\label{eq:rate_compact}
&\hspace{0cm}
\cdot\,\sum_{ij} \,\langle l| \hat{U}^\dagger \hat{F}_i \hat{U} |l'\rangle\,
\langle l'| \hat{U}^\dagger \hat{F}_j \hat{U} |l\rangle \, \tau_{ij}^{\alpha\alpha'} \quad,
\end{align}
where 
\begin{align}
\label{eq:tau}
\tau_{ij}^{\alpha\alpha'}\,=\,\text{Tr} \,\dunderline{V}_\alpha^\dagger\,\dunderline{V}_\alpha\,\dunderline{J}^i
\,\dunderline{V}_{\alpha'}^\dagger\,\dunderline{V}_{\alpha'}\,\dunderline{J}^j \quad.
\end{align} 
As expected only the combination $\dunderline{V}_\alpha^\dagger\,\dunderline{V}_\alpha$ enters into
this expression which is consistent with our discussion in Section~\ref{sec:tunneling_model} where it 
was shown that the hybridization matrices $\dunderline{\Gamma}_\alpha$ depend only on this combination, see
(\ref{eq:gamma_svd}).  

The stationary probability distribution $p_l$ in the diagonalized basis follows from
\begin{align}
\label{eq:p_golden_rule}
\sum_{l'} p_{l'}\,\Gamma_{l'\rightarrow l} \,=\,0 \quad,\quad \sum_l p_l\,=\,1\quad.
\end{align}
In an analog way one can calculate the stationary current $I_\beta$ flowing in reservoir $\beta$ from
the current rates $W^\beta_{ll'}$ in golden rule
\begin{align}
\label{eq:current_golden_rule}
I_\beta\,=\,\sum_{ll'} p_{l'} \, \Gamma^\beta_{l'\rightarrow l} \quad,
\end{align}
with
\begin{align}
\nonumber
\sum_l \Gamma^\beta_{l'\rightarrow l}\,&=\,2\pi \sum_{\alpha\alpha'} (\delta_{\alpha\beta}-\delta_{\alpha'\beta})
w(\epsilon_l-\epsilon_{l'}+\mu_\alpha - \mu_{\alpha'})\,\cdot\\
\label{eq:current_rates}
&\hspace{0cm}
\cdot\,\sum_{ij} \,\langle l| \hat{U}^\dagger \hat{F}_i \hat{U} |l'\rangle\,
\langle l'| \hat{U}^\dagger \hat{F}_j \hat{U} |l\rangle \, \tau_{ij}^{\alpha\alpha'} \quad.
\end{align}

Once the input of the matrices $\dunderline{V}_\alpha$, the coupling constants $(J_1,J_4,J_6)$ and
$(K_1,K_4,K_6)$ (determining the matrices $\dunderline{J}^i$ for $i=1,\dots,8$), the unitary transformation
$\hat{U}$ and the eigenvalues $\epsilon_l$ of the dot Hamiltonian are known, the stationary 
probabilities and the current can be calculated in a straightforward way from the above golden
rule expressions. Thereby, we have neglected small renormalizations of the dot parameters
induced by the coupling to the reservoirs which are assumed to be much smaller
than the level spacings in the dot.

\subsection{$F$-spin magnetization for two reservoirs}
\label{sec:magnetization}

We now calculate the $F$-spin magnetization of the dot
\begin{align}
\label{eq:def_magnetization}
m_F\,=\,\sqrt{\sum_{i=1}^8 (\langle \hat{F}_i \rangle)^2} 
\end{align}
for the special case of two reservoirs. We will show that the condition of zero $F$-spin 
magnetization requires special dot parameters characterizing the deviation from the fixed point model. In the 
basis of the diagonalized dot Hamiltonian the density matrix of the dot is diagonal in golden rule
approximation so that only the two diagonal generators $\hat{F}_3$ and $\hat{F}_8$ contribute to $m_F$
\begin{align}
\nonumber
m_F\,&=\,\sqrt{(\langle \hat{F}_3 \rangle)^2 + \langle \hat{F}_8 \rangle)^2}\\
\label{eq:m_F3_F8}
&=\,{1\over 2}\sqrt{(p_1-p_2)^2 + {1\over 3}(p_1+p_2-2p_3)^2}\quad.
\end{align}
Zero $F$-spin magnetization is then equivalent to an equal population of the three states
\begin{align}
\label{eq:zero_m_F}
m_F\,=\,0 \quad \Leftrightarrow \quad p_1=p_2=p_3 \quad.
\end{align}

As explained in Section~\ref{sec:tunneling_model} via (\ref{eq:diagonalization_VV}) 
the case of two reservoirs has the advantage
that both matrices $\dunderline{V}_\alpha^\dagger\,\dunderline{V}_\alpha=
\dunderline{U}_V\,\dunderline{A}_\alpha^\text{d}\,\dunderline{U}^\dagger_V$ can be diagonalized by 
a common unitary matrix $\dunderline{U}_V$ and the diagonal matrices $\dunderline{A}_\alpha^\text{d}$
are parametrized via (\ref{eq:A_decomp_N=3}) by the parameters $x_\alpha$, $p_\alpha$ and $q_\alpha$,
which fulfil the conditions (\ref{eq:x_property_two_reservoirs}-\ref{eq:q_property_two_reservoirs}) 
and (\ref{eq:p_q_condition}). Furthermore it was shown that the special property of the fixed point 
model is that the unitary transformation $\dunderline{U}_V$ can be shifted to the dot such that in
the new basis an
effective diagonal tunneling model (\ref{eq:gamma_fixed_point_two_reservoirs_diagonal}) emerges.
Thus, the particular property of the fixed point model is that the expectation value of the
$F$-spin magnetization and the current $I_\alpha$ are independent of the unitary matrix 
$\dunderline{U}_V$. In contrast, for the model away from the fixed point this is no longer the case. 

The unitary matrix $\dunderline{U}_V$ provides a mean to parametrize the dot Hamiltonian by 
convenient parameters. After transforming the dot Hamiltonian with 
$\hat{U}_V=\sum_{ll'}(\dunderline{U}_V)_{ll'}|l\rangle\langle l'|$ we take the form
\begin{align}
\label{eq:H_dot_parameters}
\hat{U}_V^\dagger H_\text{dot} \hat{U}_V \,=\,h_x\hat{F}_1 + h_y\hat{F}_2 + h_z\hat{F}_3 
+ {2\over \sqrt{3}}\Delta\hat{F}_8 \quad,
\end{align}
such that $\vec{h}$ can be interpreted as an effective magnetic field 
acting on the isospin of the up/down quark, and $\Delta$ is the level distance between the strange
quark and the average level position of the up and down quark
\begin{align}
\label{eq:Delta}
\Delta\,&=\,{1\over 2}(\epsilon_1 + \epsilon_2)\,-\,\epsilon_3 \quad,
\end{align}
see also Fig.~\ref{fig:model} for an illustration.
The eigenvalues $\epsilon_l$ of $H_\text{dot}$ and the unitary operator $\hat{U}$ can then be expressed 
by the dot parameters $\vec{h}$ and $\Delta$ by 
\begin{align}
\label{eq:epsilon_h}
\epsilon_{1/2}\,&=\,\pm {1\over 2}h + {1\over 3}\Delta \quad,\quad \epsilon_3\,=\,-{2\over 3}\Delta \quad,\\
\label{eq:U_hV}
\hat{U}\,&=\,\hat{U}_V\,\hat{U}_h\quad,\quad
\dunderline{U}_h\,=\,
\left(\begin{array}{c|c} \uline{x}_1 \,\uline{x}_2 & 0 \\ \hline 0 & 1 \end{array}\right) \quad,
\end{align}
where $h=\sqrt{h_\perp^2 + h_z^2}$, $h_\perp^2=h_x^2+h_y^2$, and
\begin{align}
\label{eq:x_12}
\uline{x}_{1/2}\,=\,{1\over\sqrt{2h(h\mp h_z)}}
\left(\begin{array}{c} \pm (h_x-ih_y) \\ h\mp h_z  \end{array}\right) \quad.
\end{align}
Inserting $\hat{U}=\hat{U}_V\hat{U}_h$ and $\dunderline{V}_\alpha^\dagger\,\dunderline{V}_\alpha=
\dunderline{U}_V\,\dunderline{A}_\alpha^\text{d}\,\dunderline{U}^\dagger_V$ in the golden rate
(\ref{eq:rate_compact}) we get
\begin{align}
\nonumber
\Gamma_{l'\rightarrow l}\,&=\,2\pi \sum_{\alpha\alpha'} w(\epsilon_l-\epsilon_{l'}+\mu_\alpha - \mu_{\alpha'})\,\cdot\\
\label{eq:rate_U_hV}
&\hspace{-1cm}
\cdot\,\sum_{ij} \,\langle l| \hat{U}^\dagger\hat{U}_V^\dagger \hat{F}_i \hat{U}_V\hat{U}_h |l'\rangle\,
\langle l'| \hat{U}^\dagger_h\hat{U}^\dagger_V \hat{F}_j \hat{U}_V\hat{U}_h |l\rangle \, \tau_{ij}^{\alpha\alpha'} \quad,
\end{align}
with 
\begin{align}
\label{eq:tau_U_V}
\tau_{ij}^{\alpha\alpha'}\,=\,\text{Tr} 
\,\dunderline{A}^d_\alpha(\dunderline{U}_V^\dagger\,\dunderline{J}^i\,\dunderline{U}_V)
\,\dunderline{A}^d_{\alpha'}(\dunderline{U}_V^\dagger\,\dunderline{J}^j\,\dunderline{U}_V) \quad.
\end{align} 
For the special case of the fixed point model where $\dunderline{J}^i={1\over 2}J\dunderline{\lambda}_i$
we can see that the unitary matrix $\dunderline{U}_V$ indeed drops out as expected due to the invariant
\begin{align}
\label{eq:invariant}
\sum_{i=1}^8 (\dunderline{U}_V^\dagger\,\dunderline{\lambda}_i\,\dunderline{U}_V)
(\hat{U}_V^\dagger \hat{F}_i \hat{U}_V) \,=\, \sum_{i=1}^8 \,\dunderline{\lambda}_i\,\hat{F}_i \quad.
\end{align}
An analog property holds for the current rate (\ref{eq:current_golden_rule}).

In the following, we consider the strong nonequilibrium regime where
the bias voltage $V=\mu_L-\mu_R>0$ is assumed to be larger than all level spacings, i.e.
\begin{align}
\label{eq:strong_noneq}
V\,>\,|h|,|\Delta \pm h/2| \quad,
\end{align}
From \eqref{eq:m_F3_F8} we see directly that the condition $m_F=0$ is equivalent to 
$\langle \hat{F}_3 \rangle = \langle \hat{F}_8 \rangle =0$. Consequently, this are two 
conditions revealing that $m_F=m_F(h_z,h_\perp,\Delta)=0$ generically defines a closed curve 
in $(h_z, h_\perp, \Delta)$-space. Inserting (\ref{eq:A_decomp_N=3}) for $\dunderline{A}_\alpha^d$, 
(\ref{eq:J_i_RG}) for $\dunderline{J}^i$, (\ref{eq:epsilon_h}) for $\epsilon_l$, and (\ref{eq:U_hV}) 
for $\hat{U}_h$, we evaluate the golden rule rates (\ref{eq:rate_U_hV}) and (\ref{eq:current_golden_rule})
in Appendix~\ref{app:golden_rule} for the special case $\dunderline{U}_V=\dunderline{\mathbbm{1}}$ 
from which we can determine the shape of this curve. This gives a generic result for the fixed point model
(where the matrix $\dunderline{U}_V$ drops out) whereas for the model away from the fixed point
we consider only the special case of a diagonal tunneling model. 

\begin{figure}
\center
 \includegraphics[width=0.5\textwidth]{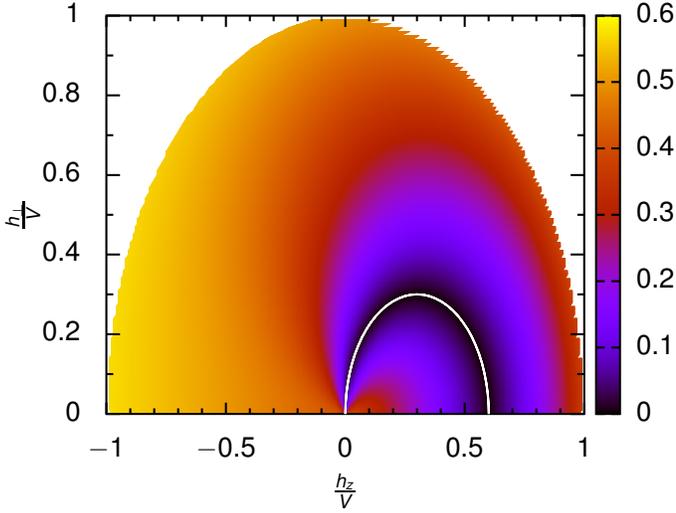}
\caption{(Color online) The $F$-spin magnetization $m_F$ in the strong nonequilibrium regime projected 
  onto the $(h_z,h_\perp)$-plane at the fixed point with $x_L=x_R=0.5$, $p_L=-p_R=0.6$, $q_L=-q_R=1.0$, 
  $J=0.0965103$, $V=10^3\, T_K$ and $\Delta=0.5 V$. The white line $h^*_\perp(h_z)$ indicates where $m_F$ is zero.}
\label{fig:magnetization}
\end{figure}

From the condition $m_F(h_z,h_\perp,\Delta)=0$ or $p_1=p_2=p_3=1/3$ we obtain in Appendix~\ref{app:golden_rule}
the two equations
\begin{align}
\label{eq:m_zero_delta}
\Delta\,&=\,x_L q_L V\,+\,{J_4^2 - J_6^2 \over J_4^2 + J_6^2} (x_L p_L V -{1\over 2} h_z) \quad,\\
\label{eq:m_zero_ellipse}
\theta_2^2 x_L^2 p_L^2 V^2\,&=\,\theta_1^2 h_\perp^2\,+\,\theta_2^2 (h_z-x_L p_L V)^2 \quad,
\end{align}
where 
\begin{align}
\label{eq:theta_1}
\theta_1^2\,&=\,J_1^2 + J_3^2 + J_{38}^2 + {1\over 2}(J_4^2 + J_6^2) \quad,\\
\label{eq:theta_2}
\theta_2^2\,&=\,2 J_1^2 + {3\over 2}J_4^2 - {1\over 2} J_6^2 \quad.
\end{align}
This means that the projection of the curve $m_F(h_z,h_\perp,\Delta)=0$ on the $(h_z,h_\perp)$-plane is 
an ellipse with the ratio 
\begin{align}
\label{eq:s_1}
s_1\,=\,\theta_1/\theta_2
\end{align}
of the two shape parameters. $\theta_1$ is the major-axis (minor-axis) if $s_1>1$ ($s_1<1$). We point 
out that this is different to the $SU(2)$-model (i.e. $J_{38}=J_4=J_6=0$) where $\theta_1$ is always the 
major-axis. Furthermore, the derivative of $\Delta$ w.r.t. $h_z$ is given by
\begin{align}
\label{eq:s_2}
s_2\,=\,{d\Delta\over d h_z}\,=\,-{1\over 2}{J_4^2 - J_6^2 \over J_4^2 + J_6^2} \quad.
\end{align}
The two parameters $s_{1/2}$ provide {\it smoking guns} for the detection of the fixed point model since
for $J_i = J/2$ and $J_{38}=0$ we obtain
\begin{align}
\label{eq:s_12_fixed_point}
s_1\,=\,1 \quad,\quad s_2\,=\,0 \quad,
\end{align}
i.e. a circle in the $(h_z,h_\perp)$-plane as shown in Fig.~\ref{fig:magnetization} and 
no dependence of $\Delta=q_L V$ on $h_z$ at the fixed point.
In this sense $1-s_1$ and $s_2$ can both be viewed as parameters measuring the distance from the 
fixed point model. Furthermore we see that the parameters $x_L p_L=-x_R p_R$ and $x_L q_L=-x_R q_R$ 
of the fixed point model can be determined from the two equations
\begin{align}
\label{eq:equations_fixed_point}
\Delta\,=\,x_L q_L V \quad,\quad h_\perp^2 + (h_z - x_L p_L V)^2 \,=\,x_L p_L^2 V^2\quad.
\end{align}
To fix the remaining parameter $x_L x_R$ and the coupling $J$ from a physical quantity 
we have also evaluated the current in Appendix~\ref{app:golden_rule} and obtained at the fixed point and
for $m_F=0$
\begin{align}
\nonumber 
I_L\,&=\,-I_R\\
\nonumber
&=\,\pi x_L x_R J^2\Big\{-{q_L - q_R \over6} \Delta - {p_L - p_R \over 4} h_z \,+ \\
\nonumber
&\hspace{1cm}
+\,{1\over 3}(4  - {q_L q_R \over 9} - {p_L p_R\over 3})V\,\Big\}\\
\nonumber
&=\,\pi J^2\Big\{{1\over 6}x_R q_R\Delta + {1\over 4}x_R p_R h_z + \\
\label{eq:current_fixed_point}
&\hspace{1cm}
{1\over 3}(4x_L x_R + {1\over 9}x_R^2 q_R^2 + {1\over 3} x_R^2 p_L^2) V \Big\}\quad,
\end{align}
where we used $x_L x_R (q_L-q_R)=-x_R q_R$ and $x_L x_R (p_L-p_R)=-x_R p_R$ in the last equation.
$J^2$ is just the overall scale of the current and the parameter $x_L x_R$ appears explicitly.
Together with $x_L+x_R=1$, the two parameters $x_{L/R}$ can thus be fixed. 

In summary, we have shown in the strong nonequilibrium regime that the condition of vanishing 
$F$-spin magnetization $m_F=0$ defines a closed curve in $(h_z, h_\perp, \Delta)$-space that is an 
ellipse in the special case of a diagonal tunneling model. A golden rule calculation has revealed 
that the geometric properties of this ellipse are a measure for the distance to the fixed point 
model where the ellipse turns into a circle being embedded in a plane defined by a constant value 
for $\Delta$. At the fixed point, the parameters of the effective model can experimentally be 
obtained from identifying the position of this circle together with measuring the current at the 
corresponding dot parameters $\vec{h}$ and $\Delta$.

\section{Summary}
\label{sec:summary}

The results obtained in this paper show that the area of nonequilibrium low-temperature transport through
generic quantum dot models contains a huge variety of interesting fixed point models not accessible in the
equilibrium case. Previous studies have analysed many generic Kondo scenarios for equilbrium systems and
used the finite voltage $V$ just as a probe for the equilibrium dot spectral density for quantum dots
coupled very asymmetrically to two reservoirs \cite{glazman_pustilnik_05}. In addition, the finite voltage
together with corresponding decay rates was just expected to act as a cutoff scale for RG flows in the
weak coupling regime \cite{rosch_etal,kehrein_etal,hs_reininghaus_09} analog to the temperature, leading
to quantitatively but {\it not} qualitatively different physical properties. In contrast, the
analysis performed in this paper shows that, for generic tunneling matrices, the cutoff set 
by the voltage is essentially different from
the temperature since it drives the system towards a fixed point characterized by a 
{\it different symmetry} compared to the equilibrium case. Our main result is that if 
an electron on a singly-occupied dot in the cotunneling regime can occupy $N$ levels flavor fluctuations 
lead to a model in the nonequilibrium situation which is essentially {\it not} $SU(N)$-invariant. In
the scaling limit for fixed values of $V$ and $T_K$, a fixed point model appears at scale $V$ where each 
reservoir is characterized by $N$ effective flavors with $N^2-1$-dimensional polarizations 
(corresponding to the $N^2-1$ generators of the
$SU(N)$-group) pointing in different directions such that the total sum is equal to zero. This leads
to a $SU(N)$-symmetric equilibrium fixed point where all reservoirs can be taken together,
but to a $SU(N)$-{\it nonsymmetric} nonequilibrium fixed point with {\it qualitatively} different 
physical properties. We have
demonstrated this for the special case $N=3$ and two reservoirs in the weak coupling regime $V\gg T_K$
and have seen that the condition of equal population of all dot states is realized for special dot 
parameters providing a {\it smoking gun} to identify the special symmetry of the nonequilibrium 
fixed point model. 

Strictly speaking the numerical solution of the RG flow shows that even for rather large ratios $D/T_K$, the
coupling constants become all equal only very close to $T_K$, where the poor man scaling approach is no
longer valid. This means that the fixed point model can not be reached for voltages $V\gg T_K$,
except for cases where the initial parameters have already been set close to the fixed point. It is therefore
of high interest for the future to develop numerically exact approaches to describe the strong coupling 
regime in nonequilibrium. In particular for voltages $V\sim T_K$ we expect that the fixed point model has
been reached and the scaling of the conductance and the $F$-spin magnetization as function of the dot 
parameters will be essentially different from the $SU(N)$-symmetric case. In agreement with 
Rfs.~\onlinecite{Moca2012,lopez_etal_13} we have demonstrated in this paper that in equilibrium the 
fixed point model is indeed reached for temperatures below the Kondo temperature $T_K$ providing 
evidence that a similiar result will also hold in the nonequilibrium case when the voltage 
reaches $T_K$. It will be interesting for the future to test this conjecture and to provide 
signatures of the nonequilibrium fixed point model in the strong coupling regime. 

Finally, it will also be very interesting for the future to study the nonequilibrium fixed points 
in regimes where the particle number of the dot is larger than one $N_\text{dot}>1$. Already
in the equilibrium case it has been demonstrated that not only the Coulomb interaction but also 
other kinds of interactions (e.g. spin-dependent terms) are very important to find the correct ground state,
see e.g. Ref.~\onlinecite{glazman_pustilnik_05} for a review. Based on this and our results for 
$N_{\text{dot}}=1$ we expect that even a richer variety of new nonequilibrium fixed point models 
has to be expected for $N_\text{dot}>1$.

\section*{Acknowledgments}
{This work was supported by the Deutsche Forschungsgemeinschaft via RTG 1995 (C.J.L. and H.S.).
We thank A.~Weichselbaum and S.-S.~B.~Lee for 
helpful discussions on the NRG setup. F.~B.~K. and J.~v.~D. acknowledge support by the Cluster of 
Excellence Nanosystems Initiative Munich; F.~B.~K. acknowledges funding from the research school IMPRS-QST.

\begin{appendix}

\section{Reservoir self energy}
\label{app:res_self_energy}

In this Appendix we calculate the greater/lesser self-energies
$\dunderline{\Sigma}^{\gtrless}_\text{res}(\omega)$ of the dot
arising from the tunneling Hamiltonian (\ref{eq:H_tun}) with an effective tunneling matrix given
by (\ref{eq:t_fixed_point}) together with the potential scattering term $V_\text{sc}$, 
see (\ref{eq:potential_scattering}). The effective d.o.s. of the reservoirs is given by unity since 
the whole nontrivial information of the reservoirs is included in the effective tunneling matrix. 
Using standard Keldysh formalism we get
\begin{align}
\label{eq:self_energy_1}
\dunderline{\Sigma}^{\gtrless}_\text{res}(\omega)\,=\,
\gamma^2\,\sum_{\alpha\alpha'}\sum_{kk'} \,\dunderline{V}_\alpha^\dagger\,
\dunderline{G}^{\gtrless}_{\alpha k,\alpha' k'}(\omega)\,\dunderline{V}_{\alpha'}\quad,
\end{align}
where $\dunderline{G}^{\gtrless}_{\alpha k,\alpha' k'}(\omega)$ are the greater/lesser reservoir Green's 
functions arising from the reservoir part of the Hamiltonian including the potential scattering term.
These Green's functions can be calculated from the Dyson equation with $V_\text{sc}$ defining the self-energy
\begin{align}
\nonumber
\dunderline{G}^{\gtrless}_{\alpha k,\alpha' k'}(\omega)\,&=\,
\dunderline{g}^{\gtrless}_{\alpha k}(\omega)\delta_{\alpha\alpha'}\delta_{kk'}\\
\nonumber
&\hspace{-1cm}
+\,v_\text{sc} \sum_{\alpha_1 k_1} \,\dunderline{g}^{\gtrless}_{\alpha k}(\omega)\,
\dunderline{V}_\alpha\,\dunderline{V}^\dagger_{\alpha_1}\,\dunderline{G}^A_{\alpha_1 k_1,\alpha' k'}(\omega)\\
\label{eq:dyson_equation_greater_lesser}
&\hspace{-1cm}
+\,v_\text{sc} \sum_{\alpha_1 k_1} \,\dunderline{g}^R_{\alpha k}(\omega)\,
\dunderline{V}_\alpha\,\dunderline{V}^\dagger_{\alpha_1}\,\dunderline{G}^{\gtrless}_{\alpha_1 k_1,\alpha' k'}(\omega)
\quad,
\end{align}
where $\dunderline{G}^A_{\alpha_1 k_1,\alpha' k'}(\omega)$ denotes the advanced Green's function which follows
from the Dyson equation
\begin{align}
\nonumber
\dunderline{G}^A_{\alpha k,\alpha' k'}(\omega)\,&=\,
\dunderline{g}^A_{\alpha k}(\omega)\delta_{\alpha\alpha'}\delta_{kk'}\\
\label{eq:dyson_equation_advanced}
&\hspace{-1cm}
+\,v_\text{sc} \sum_{\alpha_1 k_1} \,\dunderline{g}^A_{\alpha k}(\omega)\,
\dunderline{V}_\alpha\,\dunderline{V}^\dagger_{\alpha_1}\,\dunderline{G}^A_{\alpha_1 k_1,\alpha' k'}(\omega)
\quad.
\end{align}
$\dunderline{g}^x_{\alpha k}$ (with $x=R,A,\gtrless$) denote the free Green's functions of reservoir $\alpha$ 
without $V_\text{sc}$ given by
\begin{align}
\label{eq:g_R_A_res}
\dunderline{g}^{R/A}_{\alpha k}(\omega)\,&=\,{1\over \omega-\epsilon_{\alpha k}\dunderline{\mathbbm{1}}\pm i\eta}\quad,\\
\label{eq:g_<_res}
\dunderline{g}^<_{\alpha k}(\omega)\,&=\,-f_\alpha(\omega)(\dunderline{g}^R-\dunderline{g}^A)(\omega)\quad,\\
\label{eq:g_>_res}
\dunderline{g}^>_{\alpha k}(\omega)\,&=\,(1-f_\alpha(\omega))(\dunderline{g}^R-\dunderline{g}^A)(\omega)\quad.
\end{align}
Since the d.o.s. of the reservoirs is unity we get 
\begin{align}
\label{eq:g_R_A_sum_over_k}
\sum_k \dunderline{g}^{R/A}_{\alpha k}(\omega)\,&=\,\mp i\pi \dunderline{\mathbbm{1}}\quad,\\
\label{eq:g_<_sum_over_k}
\sum_k \dunderline{g}^<_{\alpha k}(\omega)\,&=\,2\pi i f_\alpha(\omega)\dunderline{\mathbbm{1}}\quad,\\
\label{eq:g_>_sum_over_k}
\sum_k \dunderline{g}^>_{\alpha k}(\omega)\,&=\,-2\pi i (1-f_\alpha(\omega))\dunderline{\mathbbm{1}}\quad.
\end{align}
Using these properties together with 
$\sum_\alpha\,\dunderline{V}_\alpha^\dagger\,\dunderline{V}_\alpha=\dunderline{\mathbbm{1}}$ and defining
\begin{align}
\label{eq:G_bar}
\dunderline{\bar{G}}^x(\omega)\,=\,\sum_{\alpha\alpha'}\sum_{kk'}\,\dunderline{V}_\alpha^\dagger\,
\dunderline{G}_{\alpha k,\alpha' k'}^x(\omega)\,\dunderline{V}_{\alpha'}\quad,
\end{align}
with $x=R,A,\gtrless$, we obtain from the Dyson equations (\ref{eq:dyson_equation_greater_lesser}) and
(\ref{eq:dyson_equation_advanced}) after a straightforward calculation
\begin{align}
\label{eq:dyson_bar_G_A}
\dunderline{\bar{G}}^A\,&=\,i\pi\dunderline{\mathbbm{1}}\,+\,i\pi v_\text{sc}\,\dunderline{\bar{G}}^A\quad,\\
\nonumber
\dunderline{\bar{G}}^<(\omega)\,&=\,-\,i\pi v_\text{sc}\,\dunderline{\bar{G}}^<(\omega) \\
\label{eq:dyson_bar_G_<}
&+\,2\pi i\sum_\alpha f_\alpha(\omega)\,
\dunderline{V}_\alpha^\dagger\,\dunderline{V}_\alpha\,\Big(\dunderline{\mathbbm{1}}
\,+\,v_\text{sc}\,\dunderline{\bar{G}}^A\Big)
\quad,\\
\nonumber
\dunderline{\bar{G}}^>(\omega)\,&=\,-\,i\pi v_\text{sc}\,\dunderline{\bar{G}}^>(\omega) \\
\label{eq:dyson_bar_G_>}
&\hspace{-1cm}
-\,2\pi i\sum_\alpha (1-f_\alpha(\omega))\,
\dunderline{V}_\alpha^\dagger\,\dunderline{V}_\alpha\,\Big(\dunderline{\mathbbm{1}}
\,+\,v_\text{sc}\,\dunderline{\bar{G}}^A\Big)
\quad.
\end{align}
Solving this set of matrix equations for $\dunderline{\bar{G}}^{\gtrless}(\omega)$ and inserting the solution in
\begin{align}
\label{eq:sigma_lesser_greater_final}
\dunderline{\Sigma}^{\gtrless}_\text{res}(\omega)\,=\,\gamma^2\,\dunderline{\bar{G}}^{\gtrless}(\omega)\quad,
\end{align}
we finally get the result (\ref{eq:self_energy_lesser}-\ref{eq:self_energy_greater}) 
for the self-energies with an effective hybridization matrix given by 
(\ref{eq:gamma_effective_dos_fixed_point}).

\section{Equilibrium ground state of the fixed point model}
\label{app:ground_state_of_fixed_point_model}

In section \ref{sec:pms_quarks}, we have argued why the dot representation has the $[3]$ 
fundamental representation while the first state of the reservoir the complex conjugate of this 
fundamental representation $[\overline{3}]$. Representing both sites by $[3]$ (or, equivalently, 
by the complex conjugate of this representation $[\overline{3}]$), leads to a decomposition of the 
Hilbert space of the composite system into a sextet and a triplet. Accordingly, a $SU(3)$-symmetric 
Hamiltonian in this representation has an either threefold or sixfold degenerate ground state which 
is in contrast to the outcome of our analysis. Choosing the complex conjugate representation 
$[\overline{3}]$ for the reservoir site instead leads to a Hilbert space that decomposes into an 
octet and a singlet. A $SU(3)$-symmetric Hamiltonian in this representation yields two different 
eigenenergies of which one is non-degenerate and the other eightfold degenerate.

We want to emphasize that this is fundamentally different to the situation in the corresponding 
$SU(2)$ model. Generally, the fundamental representation of the spin ${1 \over 2}$ [2] is equivalent 
to its complex conjugate, i.e. $[2]=[\overline{2}]$. This is consistent with the observation that no 
anti-spin ${1 \over 2}$ exists. However, this a special property of the $SU(2)$ group that holds no 
longer for $SU(N)$ with $N>2$ and we anticipate for an analog $SU(N)$-model a ground state inspired by 
flavor-antiflavor pairs.

We consider the following set of basis states for the composite system
\begin{align}
\label{eq:eigenstate_1}
\ket{u \overline{s}} \,&=\, \ket{u} \otimes \ket{\overline{s}} \quad, \\
\label{eq:eigenstate_2}
\ket{d \overline{s}} \,&=\, \ket{d} \otimes \ket{\overline{s}} \quad, \\
\label{eq:eigenstate_3}
\ket{d \overline{u}} \,&=\, \ket{d} \otimes \ket{\overline{u}} \quad, \\
\label{eq:eigenstate_4}
\ket{u \overline{d}} \,&=\, \ket{u} \otimes \ket{\overline{d}} \quad, \\
\label{eq:eigenstate_5}
\ket{s \overline{u}} \,&=\, \ket{s} \otimes \ket{\overline{u}} \quad, \\
\label{eq:eigenstate_6}
\ket{s \overline{d}} \,&=\, \ket{s} \otimes \ket{\overline{d}} \quad, \\
\label{eq:eigenstate_7}
\ket{u \overline{u}} \,&=\, \ket{u} \otimes \ket{\overline{u}} \quad, \\
\label{eq:eigenstate_8}
\ket{d \overline{d}} \,&=\, \ket{d} \otimes \ket{\overline{d}} \quad, \\
\label{eq:eigenstate_9}
\ket{s \overline{s}} \,&=\, \ket{s} \otimes \ket{\overline{s}} \quad.
\end{align}
In a quark picture, these states are meaningful since they are all eigenstates of the total charge operator
\begin{align}
\label{eq:total_charge_operator}
\hat{q}_\text{tot} \,=\, \hat{Q} + \hat{q} \quad, 
\end{align}
where $\hat{Q}=\hat{F}_3 + {1 \over \sqrt{3}} \hat{F}_8$ and $\hat{q}=\hat{f}_3 + {1 \over \sqrt{3}} \hat{f}_8$ 
are defined as usual in the quark model \cite{su3_group}, with an integer eigenvalue. This is analog to the 
observation that no elementary particle with non-integer electrical charge exist in nature.

Let the effective Hamiltonian $V_\text{eff}$ \eqref{eq:V_eff_SU3} act on the states 
(\ref{eq:eigenstate_1}-\ref{eq:eigenstate_6}), we find that 
$\ket{u \overline{s}}$, $\ket{d \overline{s}}$, $\ket{d \overline{u}}$, $\ket{u \overline{d}}$, 
$\ket{s \overline{u}}$ and $\ket{s \overline{d}}$ are eigenstates with eigenvalue $E_8={1 \over 6} J$. 
Instead, the remaining states (\ref{eq:eigenstate_7}-\ref{eq:eigenstate_9}) are no eigenstates since
\begin{align}
V_\text{eff} \ket{u \overline{u}}  \,&=\, - {J \over 3} \ket{u \overline{u}} 
- {J \over 2} \left(\ket{d \overline{d}} + \ket{s \overline{s}} \right) \quad, \\
V_\text{eff} \ket{d \overline{d}}  \,&=\, - {J \over 3} \ket{d \overline{d}} 
- {J \over 2} \left(\ket{u \overline{u}} + \ket{s \overline{s}} \right) \quad, \\
V_\text{eff} \ket{s \overline{s}}  \,&=\, - {J \over 3} \ket{s \overline{s}} 
- {J \over 2} \left(\ket{u \overline{u}} + \ket{d \overline{d}} \right) \quad.
\end{align}
Finding the remaining eigenstates is a trivial diagonalization problem in the $3 \times 3$ subspace 
of $\ket{u \overline{u}}$, $\ket{d \overline{d}}$ and $\ket{s \overline{s}}$. The first two linear combinations
\begin{align}
\ket{1} \,&=\, {1 \over \sqrt{2}} \left(\ket{u \overline{u}} - \ket{d \overline{d}}\right) \quad,\\
\ket{2} \,&=\, {1 \over \sqrt{6}} \left(\ket{u \overline{u}} + \ket{d \overline{d}} 
- 2 \ket{s \overline{s}}\right) \quad,
\end{align}
with eigenvalue $E_8$ complement the octet. Being orthogonal to $\ket{1}$ and $\ket{2}$, the singlet 
eigenstate is the ground state \eqref{eq:ground_state_fixed_point_model} with eigenvalue 
$E_\text{gs}=-{4 \over 3} J$. We note that this set of eigenstates is the same as for pseudoscalar 
mesons in the light quark model\cite{quarks}.

\section{Evaluation of golden rule rate}
\label{app:golden_rule}

In this appendix we evaluate the golden rule rates (\ref{eq:rate_U_hV}) and (\ref{eq:current_golden_rule})
for the special case $\dunderline{U}_V=\dunderline{\mathbbm{1}}$. 
We denote the three states by the quark flavors, i.e. $l=1,2,3\equiv u,d,s$. 
First, we evaluate the matrix elements $\tau_{i j}^{\alpha \alpha^\prime}$ from (\ref{eq:tau_U_V}) 
by employing the algebra of the Gell-Mann matrices. Writing
\begin{align}
\label{eq:tau_bar}
\tau_{i j}^{\alpha \alpha^\prime}\,=\,x_\alpha x_{\alpha'} \bar{\tau}_{i j}^{\alpha \alpha^\prime}\quad,
\end{align}
we obtain for the non-vanishing matrix elements
\begin{align}
\label{eq:tau_bar_11}
\bar{\tau}_{1 1}^{\alpha \alpha^\prime} &=\, \bar{\tau}_{2 2}^{\alpha \alpha^\prime} =\,  
2 J^2_1 M^-_{1,\alpha \alpha^\prime} \quad, \\
\label{eq:tau_bar_12}
\bar{\tau}_{1 2}^{\alpha \alpha^\prime} &=\, -\bar{\tau}_{2 1}^{\alpha \alpha^\prime} 
=\, 2 i J^2_1 M^-_{2,\alpha \alpha^\prime}  \quad, \\
\label{eq:tau_bar_44}
\bar{\tau}_{4 4}^{\alpha \alpha^\prime} &=\, \bar{\tau}_{5 5}^{\alpha \alpha^\prime} 
=\, J^2_4 M^{++}_{3,\alpha \alpha^\prime}  \quad, \\
\label{eq:tau_bar_45}
\bar{\tau}_{4 5}^{\alpha \alpha^\prime} &=\, - \bar{\tau}_{5 4}^{\alpha \alpha^\prime} 
=\, i J^2_4 M^{+-}_{3,\alpha \alpha^\prime}  \quad, \\
\label{eq:tau_bar_66}
\bar{\tau}_{6 6}^{\alpha \alpha^\prime} &=\, \bar{\tau}_{7 7}^{\alpha \alpha^\prime} 
=\, J^2_6 M^{-+}_{3,\alpha \alpha^\prime}  \quad, \\
\label{eq:tau_bar_67}
\bar{\tau}_{6 7}^{\alpha \alpha^\prime} &=\, - \bar{\tau}_{7 6}^{\alpha \alpha^\prime} 
=\, i J^2_6 M^{--}_{3,\alpha \alpha^\prime}  \quad, \\
\nonumber
\bar{\tau}^{\alpha \alpha^\prime}_{i j} \Huge|_{i,j \in (3,8)} 
&=\, 2J_{3i} J_{3j} M^+_{1,\alpha \alpha^\prime} 
+ 2 J_{i 8} J_{j 8} M_{4,\alpha \alpha^\prime} \\
\label{eq:tau_bar_ij_in_38}
&\hspace{0cm}
+ {2 \over \sqrt{3}} \left(J_{3 i} J_{j 8} + J_{i 8} J_{3 j} \right) M^+_{2,\alpha \alpha^\prime} \quad,
\end{align}
where $J_{33}=J_3$, $J_{88}=J_8$, and
\begin{align}
\label{eq:M_1}
M^\sigma_{1,\alpha \alpha^\prime} &=\, \overline{q}_\alpha \overline{q}_{\alpha^\prime} 
+ \sigma p_\alpha p_{\alpha^\prime} \quad,\\
\label{eq:M_2}
M^\sigma_{2,\alpha \alpha^\prime} &=\, p_\alpha \overline{q}_{\alpha^\prime} 
+ \sigma \overline{q}_\alpha p_{\alpha^\prime} \quad, \\
\label{eq:M_3}
M^{\sigma \sigma^\prime}_{3,\alpha \alpha^\prime} &=\, \left(\overline{q}_\alpha  
+ \sigma p_\alpha\right) \tilde{q}_{\alpha^\prime} + \sigma^\prime \tilde{q}_\alpha \left(\overline{q}_{\alpha^\prime} 
+ \sigma p_{\alpha^\prime} \right) \quad, \\
\label{eq:M_4}
M_{4,\alpha \alpha^\prime} &=\, 1+ {p_\alpha p_{\alpha^\prime} 
+ q_\alpha q_{\alpha^\prime} \over 3} - {q_\alpha + q_{¸\alpha^\prime} \over 3} \quad,
\end{align}
with $\overline{q}_\alpha=1+{q_\alpha \over 3}$ and $\tilde{q}_\alpha=1 - {2 q_\alpha \over 3}$. 
Introducing the notation
\begin{align}
\label{eq:chi_13}
\chi_{1/3}^{\alpha \alpha^\prime} &= {\pi \over 2} 
\left(\tau_{1 1}^{\alpha \alpha^\prime} \pm \tau_{3 3}^{\alpha \alpha^\prime}\right) \quad,\\
\label{eq:chi_2s}
\chi_{2}^{\alpha \alpha^\prime} &= i \pi \tau_{1 2}^{\alpha \alpha^\prime} \quad,\quad
\chi_{s}^{\alpha} = x_\alpha \tilde{q}_\alpha \quad,\\
\label{eq:chi_ud}
\chi_{u/d}^{\alpha} &= 2 \pi x_\alpha \left[ J^2_+
\left(\overline{q}_\alpha \pm p_\alpha \phi_z \right)
+ J^2_- \left(p_\alpha  \pm \overline{q}_\alpha \phi_z \right) \right] \quad, 
\end{align}
with $J^2_\pm={1 \over 2}\left(J^2_4 \pm J^2_6\right)$ and $\phi_z= {h_z \over h}$, we obtain
by inserting \eqref{eq:U_hV} and \eqref{eq:tau_bar} in \eqref{eq:rate_U_hV} after
a straightforward calculation
\begin{align}
\nonumber
\Gamma_{d\rightarrow u} &=\, \sum_{\alpha \alpha^\prime} w(\mu_{\alpha} - \mu_{\alpha^\prime} - h)\,\cdot \\
\label{eq:rate_du}
&\hspace{1cm} 
\cdot\,\left[\chi_{1}^{\alpha \alpha^\prime} 
- \chi_2^{\alpha \alpha^\prime} \phi_z + \chi_3^{\alpha \alpha^\prime} \phi_z^2\right]  \quad,\\
\nonumber
\Gamma_{u\rightarrow d} &=\, \sum_{\alpha \alpha^\prime} w(\mu_{\alpha} - \mu_{\alpha^\prime} + h)\,\cdot  \\
\label{eq:rate_ud}
&\hspace{1cm} 
\cdot\,\left[\chi_{1}^{\alpha \alpha^\prime} 
+ \chi_2^{\alpha \alpha^\prime} \phi_z + \chi_3^{\alpha \alpha^\prime} \phi_z^2\right] \quad,\\
\label{eq:rate_su}
\Gamma_{s\rightarrow u} &=\, \sum_{\alpha \alpha^\prime} 
w(\mu_{\alpha} - \mu_{\alpha^\prime} - \Delta - {h \over 2}) 
\chi_u^\alpha \chi_s^{\alpha^\prime} \quad,\\
\label{eq:rate_us}
\Gamma_{u\rightarrow s} &=\, \sum_{\alpha \alpha^\prime} 
w(\mu_{\alpha} - \mu_{\alpha^\prime} + \Delta + {h \over 2})
\chi_s^\alpha \chi_u^{\alpha^\prime} \quad,\\
\label{eq:rate_sd}
\Gamma_{s\rightarrow d} &=\, \sum_{\alpha \alpha^\prime} 
w(\mu_{\alpha} - \mu_{\alpha^\prime} - \Delta + {h \over 2}) 
\chi_d^\alpha \chi_s^{\alpha^\prime} \quad,\\
\label{eq:rate_ds}
\Gamma_{d\rightarrow s} &=\, \sum_{\alpha \alpha^\prime} 
w(\mu_{\alpha} - \mu_{\alpha^\prime} + \Delta - {h \over 2}) 
\chi_s^\alpha \chi_d^{\alpha^\prime} \quad.
\end{align}

In the following we consider the case of two reservoirs in the strong nonequilibrium regime as 
defined in \eqref{eq:strong_noneq}. From the properties 
(\ref{eq:x_property_two_reservoirs}-\ref{eq:q_property_two_reservoirs}) 
and the results (\ref{eq:tau_bar}-\ref{eq:tau_bar_ij_in_38}) for $\tau_{ij}^{\alpha \alpha^\prime}$, we obtain
\begin{widetext}
\begin{align}
\label{eq:rate_du_two_res}
\Gamma_{d\rightarrow u} &=\,\left[\chi_{1}^{LR} - \chi_2^{LR} \phi_z + \chi_3^{LR} \phi_z^2\right] (V-h)  
+ w(-h) \sum_\alpha \left[\chi_{1}^{\alpha \alpha} + \chi_3^{\alpha \alpha} \phi_z^2\right] \quad, \\
\label{eq:rate_ud_two_res}
\Gamma_{u\rightarrow d} &=\, \Gamma_{d\rightarrow u} + 2 \chi_{2}^{LR} \phi_z V 
+ h \left(\chi_{1} + \chi_{3} \phi_z^2 \right) \quad,  \\
\label{eq:rate_su_two_res}
\Gamma_{s\rightarrow u} &=\, \chi_u^L \chi_s^R \left(V - \Delta - {h \over 2} \right) 
+ w(-\Delta - {h \over 2}) \sum_\alpha \chi_u^\alpha \chi_s^\alpha \quad, \\
\label{eq:rate_us_two_res}
\Gamma_{u\rightarrow s} &=\, \Gamma_{s\rightarrow u} + \left(\chi_s^L \chi_u^R 
- \chi_u^L \chi_s^R \right) V + \left(\Delta + {h \over 2} \right) \chi_u \chi_s \quad,  \\  
\label{eq:rate_sd_two_res}
\Gamma_{s\rightarrow d} &=\, \chi_d^L \chi_s^R \left(V - \Delta + {h \over 2} \right)  
+ w(-\Delta + {h \over 2}) \sum_\alpha \chi_d^\alpha \chi_s^\alpha \quad, \\
\label{eq:rate_ds_two_res}
\Gamma_{d\rightarrow s} &=\, \Gamma_{s\rightarrow d} + \left(\chi_s^L \chi_d^R 
- \chi_d^L \chi_s^R \right) V + \left(\Delta - {h \over 2} \right) \chi_d \chi_s \quad.
\end{align}
\end{widetext}
where we have defined
\begin{align}
\label{eq:chi_13_mean}
\chi_{1/3} \,&=\, \sum_{\alpha \alpha^\prime} \chi_{1/3}^{\alpha \alpha^\prime}\,=\, 
\pi \left[J_1^2 \pm J_3^2 \pm  J_{38}^2\right] \quad, \\
\label{eq:chi_ud_mean}
\chi_{u/d} \,&=\, \sum_\alpha \chi_{u/d}^\alpha \,=\, 2 \pi (J^2_+ \pm J^2_- \phi_z) \quad, \\
\label{eq:chi_s_mean}
\chi_s \,&=\, \sum_\alpha \chi_s^\alpha \,=\, 1 \quad,
\end{align}
and note that 
\begin{align}
\label{eq:chi_2_LR}
\chi_2^{LR} &= - 2 \pi J_1^2 x_L p_L \quad, \\
\nonumber
\chi_s^L \chi_{u/d}^R - \chi_{u/d}^L \chi_s^R &=\, 
- 2 \pi \left[x_L q_L \left(J^2_+ \pm J^2_- \phi_z\right)  \right.\\
\label{eq:difference_chi_sud}
&\hspace{0.5cm}
\left. + x_L p_L \left(J^2_- \pm J^2_+ \phi_z\right) \right] \quad.
\end{align}
The stationary probability distribution $p_l$ follows from inserting 
(\ref{eq:rate_du_two_res}-\ref{eq:rate_ds_two_res}) in (\ref{eq:p_golden_rule}). 
Finally, we can compute $m_F$ from \eqref{eq:m_F3_F8}.

We note that $m_F=0$ is equivalent to $\langle \hat{F}_3 \rangle = \langle \hat{F}_8 \rangle =0$. 
Therefore, we consider $\langle \hat{F}_3 \rangle = {1 \over 2} (p_u - p_d)$ 
and $\langle \hat{F}_8 \rangle = {1 \over 3} (p_u + p_d - 2 p_s)$ in the following and analyze under 
which conditions both expectation values become zero in the strong nonequilibrium regime. 
A cumbersome but straightforward analysis yields
\begin{align}
\nonumber
\left\langle \hat{F}_3 \right\rangle \,&=\, 
{1 \over 2 N} \left[\mathcal{F}_1 (\Gamma_{s\rightarrow u} + \Gamma_{s\rightarrow d})\right.\\
\label{eq:average_F3}
&\hspace{1cm}
\left. + \mathcal{F}_2 (\Gamma_{s\rightarrow u} - \Gamma_{s\rightarrow d})\right] \quad, \\
\nonumber
{2\over \sqrt{3}}\left\langle \hat{F}_8 \right\rangle \,&=\,
-{1 \over 3 N} \left\lbrace \mathcal{F}_1\left[2 \pi J^2_+ \left(h - 2 x_L p_L \phi_z V \right)
\right.\right.\\
\nonumber
&\hspace{0cm}
\left.\left.+ 4 \pi J^2_- (\Delta - x_L q_L V) \phi_z + \Gamma_{s\rightarrow u} - \Gamma_{s\rightarrow d} \right]
\right. \\
\label{eq:average_F8}
&\hspace{-1cm}
\left. + \mathcal{F}_2 \left[2 \mathcal{F}_2 
+ 2  \left(\Gamma_{d\rightarrow u} + \Gamma_{u\rightarrow d}\right)
+ \Gamma_{s\rightarrow u} + \Gamma_{s\rightarrow d} \right]\right\rbrace \quad.
\end{align}
Here, the factor $N$ follows from the normalization condition \eqref{eq:p_golden_rule}. 
Furthermore, we have defined the following functions in $(h_z,h_\perp,\Delta)$-space
\begin{align}
\nonumber
\mathcal{F}_1 &=\, - 2 \chi_{2}^{LR} \phi_z V - h \left(\chi_{1} + \chi_{3} \phi_z^2 \right) \\
\label{eq:function_F1}
&\hspace{0cm}
- \pi \left[ J^2_+ \left(h - 2 x_L p_L \phi_z V \right) 
+ 2 J^2_- \left(\Delta - x_L q_L V\right) \phi_z \right] \quad, \\
\label{eq:functionf_F2}
\mathcal{F}_2 &=\, \pi \left[2 J^2_+ \left(\Delta - x_L q_L V \right)
+ J^2_- \left(h_z - 2 p_l V\right) \right] 
\end{align}
$\mathcal{F}_1=\mathcal{F}_2=0$ fulfills the condition 
$\langle \hat{F}_3 \rangle = \langle \hat{F}_8 \rangle =0$. Moreover, it defines a curve in 
$(h_z,h_\perp,\Delta)$-space that provides us with a tool to measure the distance to the fixed 
point model. $\mathcal{F}_2=0$ directly yields \eqref{eq:m_zero_delta} and defines the plane in 
$(h_z,h_\perp,\Delta)$-space where the curve lies in. The shape of the curve follows from $\mathcal{F}_1=0$. 
To that end, we insert \eqref{eq:m_zero_delta} into \eqref{eq:function_F1} and 
obtain \eqref{eq:m_zero_ellipse}. That is, we project the curve onto the $(h_z,h_\perp)$-plane.

Finally, we prove \eqref{eq:current_fixed_point}. To this end, we decompose \eqref{eq:current_golden_rule} as
\begin{align}
\notag
\left\langle I_\beta \right\rangle &=\, 
\sum_{l l^\prime} \Gamma^\beta_{l^\prime\rightarrow l} p_{l^\prime} \\
\label{eq:current_decomposition}
&= \, I^\beta_0 + I^\beta_3 \left\langle \hat{F}_3 \right\rangle 
+ I^\beta_8 {2\over\sqrt{3}}\left\langle \hat{F}_8 \right\rangle \quad,
\end{align}
with
\begin{align}
\label{eq:I_0}
I^\beta_0 &=\, {1 \over 3}  \sum_{l l^\prime} \Gamma^\beta_{l^\prime\rightarrow l} \quad, \\
\label{eq:I_3}
I^\beta_3 &= \sum_l \left(\Gamma^\beta_{u\rightarrow l} - \Gamma^\beta_{d\rightarrow l}\right) \quad, \\
\label{eq:I_8}
I^\beta_8 &= {1 \over 2} \sum_l \left(\Gamma^\beta_{u\rightarrow l} 
+ \Gamma^\beta_{d\rightarrow l} - 2 \Gamma^\beta_{s\rightarrow l} \right) \quad.
\end{align}
Evaluating \eqref{eq:current_rates} for two reservoirs in the strong nonequilibrium regime 
\eqref{eq:strong_noneq}, we can express (\ref{eq:I_0}-\ref{eq:I_8}) in terms of $\bar{\tau}^{\alpha \alpha^\prime}_{ij}$
\begin{widetext}
\begin{align}
\nonumber
I^L_0 &=\, {\pi \over 3} x_L x_R \Big\lbrace \big[2 \bar{\tau}^{LR}_{11} 
+ \bar{\tau}^{LR}_{33} + 2 \left(\bar{\tau}^{LR}_{44} + \bar{\tau}^{LR}_{66} \right) 
+ \bar{\tau}^{LR}_{88} \big] V \\
\label{eq:I_0_two_res}
& \hspace{2cm}
+ 2 i\left(\bar{\tau}^{LR}_{45} + \bar{\tau}^{LR}_{67} \right) \Delta 
+ i\left(2 \bar{\tau}^{LR}_{12} + \bar{\tau}^{LR}_{45} - \bar{\tau}^{LR}_{67} \right) h_z \Big\rbrace \quad, \\     
\nonumber
I^L_3 &=\, \pi x_L x_R \Big\lbrace \big[2 i \bar{\tau}^{LR}_{12} 
+ {2 \over \sqrt{3}} \bar{\tau}^{LR}_{38} + \bar{\tau}^{LR}_{44} 
- \bar{\tau}^{LR}_{66} + i \left(\bar{\tau}^{LR}_{45} - \bar{\tau}^{LR}_{67}\right) \big] \phi_z V  \\
\nonumber
& \hspace{2cm}
+ \big[\bar{\tau}^{LR}_{44} - \bar{\tau}^{LR}_{66} + i\left(\bar{\tau}^{LR}_{45} 
- \bar{\tau}^{LR}_{67}\right)\big] \phi_z \Delta \\
\label{eq:I_3_two_res}
& \hspace{2cm}
+ \big[\bar{\tau}^{LR}_{11} + \bar{\tau}^{LR}_{33} + \left(\bar{\tau}^{LR}_{11} 
- \bar{\tau}^{LR}_{33} \right) \phi_z^2 + {1 \over 2} \left(\bar{\tau}^{LR}_{44} + \bar{\tau}^{LR}_{66} \right) 
+ {i \over 2} \left(\bar{\tau}^{LR}_{45} + \bar{\tau}^{LR}_{67} \right) \big] h \Big\rbrace\quad, \\
\nonumber
I^L_8 &=\, {\pi \over 2} x_L x_R \Big\lbrace \big[2 \bar{\tau}^{LR}_{11} + \bar{\tau}^{LR}_{33} 
- \left(\bar{\tau}^{LR}_{44} + \bar{\tau}^{LR}_{66} + \bar{\tau}^{LR}_{88}\right) 
+ 3i \left(\bar{\tau}^{LR}_{45} + \bar{\tau}^{LR}_{67} \right) \big] V \\
\nonumber
& \hspace{2cm}
+ \big[3 \left(\bar{\tau}^{LR}_{44} + \bar{\tau}^{LR}_{66} \right) 
- i \left(\bar{\tau}^{LR}_{45} + \bar{\tau}^{LR}_{67} \right) \big] \Delta \\
\label{eq:I_8_two_res}
& \hspace{2cm}
+ \big[2 i \bar{\tau}^{LR}_{12} + {3 \over 2} \left(\bar{\tau}^{LR}_{44} 
- \bar{\tau}^{LR}_{66} \right) - {i \over 2} \left(\bar{\tau}^{LR}_{45} 
- \bar{\tau}^{LR}_{67}\right) \big] h_z \Big\rbrace \quad.          
\end{align}
If we consider $m_F=0$, the current $I^\beta$ is completely equal to $I^\beta_0$. Therefore, 
we can evaluate \eqref{eq:I_3_two_res} using (\ref{eq:tau_bar_11}-\ref{eq:tau_bar_ij_in_38}) 
at the fixed point and obtain \eqref{eq:current_fixed_point}.
\end{widetext}

\end{appendix}

\end{document}